\documentclass[superscriptaddress,floatfix,table,pla,3p,twocolumn]{elsarticle}
\usepackage{graphicx}
\usepackage{amsmath}
\usepackage{amssymb}
\usepackage{amsfonts}
\usepackage{tcolorbox}
\usepackage{color}
\tcbuselibrary{skins}
\tcbuselibrary{theorems} 
\usepackage{dsfont}
\usepackage{soul}
\usepackage{hyperref}
\usepackage{physics}

\begin{document}

\title{{Thermodynamics of relativistic quantum fields confined in cavities}}

\author[1,2,3,4]{David Edward Bruschi\corref{cor1}}
\ead{david.edward.bruschi@gmail.com}
\author[2,5]{Benjamin Morris}
\ead{benjamin.morris@nottingham.ac.uk}
\author[5,6]{Ivette Fuentes}
\ead{ivette.fuentes@nottingham.ac.uk}
\cortext[cor1]{Corresponding author}
\address[1]{Theoretical Physics, Universit\"at des Saarlandes, 66123 Saarbr\"ucken, Germany}
\address[2]{York Centre for Quantum Technologies, Department of Physics, University of York, Heslington, YO10 5DD York, UK}
\address[3]{Racah Institute of Physics and Quantum Information Science Centre, the Hebrew University of Jerusalem, Givat Ram, 91904 Jerusalem, Israel}
\address[4]{Central European Institute of Technology (CEITEC), Brno University of Technology, 621 00 Brno, Czech Republic}
\address[5]{School of Mathematical Sciences, University of Nottingham, University Park, Nottingham NG7 2RD, United Kingdom}
\address[6]{Faculty of Physics, University of Vienna, Boltzmanngasse 5, A-1090 Vienna, Austria}

\begin{abstract}
We investigate the quantum thermodynamical properties of localised relativistic quantum fields, and how they can be used as quantum thermal machines. We study the efficiency and power of energy transfer between the classical \textit{gravitational} degrees of freedom, such as the energy input due to the motion of boundaries or an impinging gravitational wave, and the excitations of a confined quantum field. We find that the efficiency of energy transfer depends dramatically on the input initial state of the system. Furthermore, we investigate the ability of the system to extract energy from a gravitational wave and store it in a \textit{battery}. This process is inefficient in optical cavities but is significantly enhanced when employing trapped Bose Einstein condensates. We also employ standard fluctuation results to obtain the work probability distribution, which allows us to understand how the efficiency is related to the dissipation of work. Finally, we apply our techniques to a setup where an impinging gravitational wave excites the phononic modes of a Bose Einstein condensate. We find that, in this case, the percentage of energy transferred to the phonons approaches unity after a suitable amount of time. {These results give a quantitative insight into the thermodynamic behaviour of relativistic quantum fields confined in cavities.} 
\end{abstract}
\maketitle

\newpage
\section*{Introduction}
The quest for quantum technologies, such as quantum computers and quantum sensors, is leading technological and scientific revolutions in many areas of physics. This pursuit has stimulated research in many novel scientific directions in the past few decades, such as quantum computing \cite{Ladd:Jelezko:2010}, quantum cryptography and quantum key distribution \cite{Gisin:Ribordy:2002}, to name a few. 
Furthermore, recent breakthroughs in the understanding of the thermodynamics of systems that operate in quantum regimes have opened the door to novel theoretical and experimental developments of quantum thermal machines, such as single atom heat engines \cite{rossnagel2016single} and quantum refrigerators \cite{Correa:Palao:2014}, whose behaviour can challenge our understanding of thermodynamics  \cite{Goold:Huber:2016}. 
Typical regimes where such technologies are expected to operate are well within the realm of quantum mechanics, which is normally sufficient to describe phenomena that occur in the microscopic world.
However, in the past decade it has been shown that relativity can also enter the game. 

It is commonly believed that relativity concern mostly, if not solely, large scales, such as those where satellites operate or cosmological ones. However, the past few years have witnessed a growing body of work aimed at investigating the effects of relativity, such as high velocities and curvature, on localised relativistic quantum fields to be employed for quantum information processing \cite{Alsing:Fuentes:2012}. For example, it was shown that, in general, relativistic motion of cavities that contain quantum fields, such as the electromagnetic field, can be used to generate specific multimode entangled states with the aim of exploiting them for quantum information processing \cite{Bruschi:Fuentes:2012,Friis:Bruschi:2012}. Furthermore, it was shown that micrometer quantum systems, known as Bose-Einstein condensates (BECs), can in principle be used to detect gravitational waves emitted in (kHz) frequency domains typical of binary neutron star mergers and pulsars \cite{Sabin:Bruschi:2014}. Interestingly, the predictions of this body of work are now starting to be tested in the laboratory \cite{SandboChang:Simoen:2017}. 
In general, the results of this line of research lie at the intersection of relativity and quantum theory. In order for these systems to be employed in future technologies, such as generators of resources for quantum computing or detectors for gravitational waves, it is necessary to answer the following fundamental question: \textit{what is the quantum thermodynamical performance of such relativistic and quantum ``machines''?} 

In this work we investigate the quantum thermodynamical performance of ideal localised relativistic quantum fields to be used as the core constituents of relativistic quantum machines. We focus our attention on quantum fields confined within cavities that can be affected by motion or gravity, for example in the form of moving boundary conditions \cite{Lindkvist:Sabin:2014} or impinging gravitational waves \cite{Sabin:Bruschi:2014}. Such systems have undergone extensive study in the past few years  \cite{Alsing:Fuentes:2012}. However, in this work we take the novel approach of employing notions from quantum thermodynamics in order to understand their performance as thermal machines.

We also focus on scenarios where the initial states belong to the class of Gaussian states, which are readily produceable and accessible in many quantum optics laboratories. Since the transformations that occur are linear, this allows us to employ powerful mathematical techniques from Continuous Variables \cite{Adesso:Ragy:2014} to compute all relevant quantities of interest, such as changes of entropies and energies. 
We are able to compute the efficiency of these machines in converting the classical work provided, for example the energy necessary to accelerate a cavity or the energy that a gravitational wave transfers to the field excitations, into quantum excitations of the field that{, in the future,} can be experimentally accessed and manipulated. Surprisingly, we find that the efficiency is greatly dependent on the initial state of the field and that two very different scenarios can occur. For initial thermal and single-mode-squeezed states, the efficiency is completely independent of the magnitude of the changes induced by the classical external source. For initial beam splitted thermal states and two-mode-squeezed states, the efficiency depends on the magnitude of these changes. We believe that these results underline, once more, the importance of quantum correlations in quantum setups.

In order to establish more firmly our results, we choose a well-characterised interaction Hamiltonian and provide a protocol for the extraction of the available energy from a cavity into a ``battery'' \cite{Goold:Huber:2016,Mitchison:Huber:2016}.  We show that energy extraction is not substantial for optical cavities, however, a much higher amount of energy can be harvested in cavities containing BECs. We leave it to future work to find more suitable protocols aimed at higher degrees of efficiency in energy storage.

Finally, we specialise to a setup of a trapped BEC which is subject to an impinging gravitational wave. The spacetime within the BEC trap is affected, or ``stretched and compressed'' \cite{Sabin:Bruschi:2014} changing the state of the phononic field \cite{Sabin:Bruschi:2014}. In this context, we compute the efficiency of energy transfer between the wave and the available phononic modes, whose frequencies are in resonance with the drive. We find that this efficiency increases with time and reaches a constant value, close to unity within ultra-cold experimental setups. {These results give an insight into promising future possibilities of harvesting energy from classical relativistic degrees of freedom using confined quantum fields.} 

We conclude by noting that our results are general and can be applied to any scenario where Bogoliubov transformations are present in the dynamics of quantum fields. For example, this can also occur when localised excitations are considered in scenarios such as an expanding universe \cite{Ball:Fuentes-Schuller:2006} or the evolution of stars into black holes \cite{Hawking:1974}. Our techniques show that localised quantum fields are promising candidates as the core framework for future relativistic quantum technologies.

The paper is organised as follows. In Section \ref{background} we introduce the necessary tools for this work. In Section \ref{performance} we compute the performance of the cavities in converting the classical energy into quantum excitations. In Section \ref{section:role:initial:state} we discuss the role of the initial state in the harvesting of energy. In Section \ref{charging:battery} we provide a simple protocol aimed at extracting the energy from the accessible modes of the cavity and storing it in a battery. In section \ref{gravitational:wave:application} we provide an application of our techniques to a scenario where a gravitational wave impinges on a BEC confined within a trap. Finally, in Section \ref{conclusions} we provide concluding remarks on the current status of the art and open directions.

Throughout this work we employ $(-,+,+,+)$ as signature for the metric, lowercase bold font stands for matrices and occasionally for vectors in $3$-dimensions.

\section{Background}\label{background}

\subsection{Quantum field theory}
In this work particles are excitations of an uncharged scalar massless quantum field $\hat \Phi$, which well approximates one polarisation of the electromagnetic field \cite{Friis:Lee:2013} and a phononic field of a BEC. The field $\hat \Phi(t,\boldsymbol{x})$ is in general defined on a curved $3+1$ dimensional spacetime with coordinates $(t,\boldsymbol{x})$ and metric $g_{\mu\nu}$. The scalar field $\hat \Phi$ satisfies the Klein-Gordon equation $\square\hat \Phi=0$, where $\square:=(\sqrt{-g})^{-1}\partial_{\mu}\sqrt{-g}\,\partial^{\mu}$ is the D'Alambertian operator and $g$ is the determinant of the metric.
We are interested in localised fields and we choose to confine the field $\hat \Phi$ in a cavity of size $\boldsymbol{L}=(L_x,L_y,L_z)$. The spectrum of solutions $\{\phi_{\boldsymbol{n}}\}$ to the Klein-Gordon equation is therefore discrete and $\boldsymbol{n}=(n,m,p)$, where $n,m,p\in\mathbb{N}$. In case the spacetime has an (asymptotic) time-like Killing vector $\partial_{\tau}$, it is convenient to find an orthonormal complete set of mode solutions $\{\phi_{\boldsymbol{n}}\}$ to the Klein-Gordon equation which satisfy the eigenvalue equation $\partial_{\tau}\,\phi_n=i\,\omega_{\boldsymbol{n}}\,\phi_{\boldsymbol{n}}$, where $\omega_{\boldsymbol{n}}$ is the eigenvalue, which we will identify with the frequency of the excitations.

The field $\hat \Phi$ can be expanded in terms of the mode solutions $\{\phi_{\boldsymbol{n}}\}$ as $\hat \Phi=\sum_{\boldsymbol{n}}[\hat a_{\boldsymbol{n}}\,\phi_{\boldsymbol{n}}+\hat a^{\dag}_{\boldsymbol{n}}\,\phi^*_{\boldsymbol{n}}]$, where the creation and annihilation operators $\hat a_{\boldsymbol{n}},\hat a^{\dag}_{\boldsymbol{n}}$ satisfy the canonical commutation relations $[\hat a_{\boldsymbol{n}},\hat a^{\dag}_{\boldsymbol{n}'}]=\delta_{\boldsymbol{n},\boldsymbol{n}'}$, and all other commutators vanish. The annihilation operators $a_{\boldsymbol{n}}$ define the vacuum state $|0\rangle$ through the condition $\hat a_{\boldsymbol{n}}\,|0\rangle=0$ for all $\boldsymbol{n}$. In general, one might choose a different set of solutions $\{\tilde{\phi}_{\boldsymbol{n}}\}$ to the Klein-Gordon equation. This can occur, for example, when there are two inequivalent (asymptotic) Killing vectors i.e., $\partial_{\tau}$ and $\partial_{\tilde{\tau}}$. Therefore, in addition to the eigenvalue equation $\partial_{\tau}\,\phi_n=i\,\omega_{\boldsymbol{n}}\,\phi_{\boldsymbol{n}}$, one can choose a set of modes $\{\tilde{\phi}_{\boldsymbol{n}}\}$ that satisfies the eigenvalue equation $\partial_{\tilde{\tau}}\,\tilde{\phi}_{\boldsymbol{n}}=i\,\tilde{\omega}_{\boldsymbol{n}}\,\tilde{\phi}_{\boldsymbol{n}}$. The field can also be expanded in terms of the second set of solutions $\{\tilde{\phi}_{\boldsymbol{n}}\}$ as $\hat \Phi=\sum_{\boldsymbol{n}}[\hat b_{\boldsymbol{n}}\,\tilde{\phi}_{\boldsymbol{n}}+\hat b^\dag_{\boldsymbol{n}}\,\tilde{\phi}^*_{\boldsymbol{n}}]$, where the creation and annihilation operators $\hat b^\dag_{\boldsymbol{n}},\hat b_{\boldsymbol{n}}$ satisfy the canonical commutation relations $[\hat b_{\boldsymbol{n}},\hat b^{\dag}_{\boldsymbol{n}'}]=\delta_{\boldsymbol{n},\boldsymbol{n}'}$, and all other commutators vanish. The new annihilation operators $\hat b_{\boldsymbol{n}}$ define the vacuum $|\tilde{0}\rangle $ through the condition $\hat b_{\boldsymbol{n}}\,|\tilde{0}\rangle=0$ for all $\boldsymbol{n}$. Notice that, since the operators $\hat a_{\boldsymbol{n}}$ and $\hat b_{\boldsymbol{n}}$ are inequivalent, this implies also that the vacua $|0\rangle $ and $|\tilde{0}\rangle $ are inequivalent i.e., $|0\rangle\neq|\tilde{0}\rangle $. The inequivalence of these operators, which translates into the fact that there is no unique definition of particle in quantum field theory, is at the core of all the great predictions within this area of physics, such as the Unruh effect \cite{Unruh:1976}, the Hawking effect \cite{Hawking:1974}, the creation of particles due to the expansion of the universe \cite{Birrell:Davies:1984} and the dynamical Casimir effect \cite{Dodonov:2010}.

The two sets of operators $\{a_{\boldsymbol{n}},a^{\dag}_{\boldsymbol{n}}\}$ and $\{\hat b_{\boldsymbol{n}},\hat b^\dag_{\boldsymbol{n}}\}$, or equivalently the sets of mode solutions $\{\phi_{\boldsymbol{n}}\}$ and $\{\tilde{\phi}_{\boldsymbol{n}}\}$, are related by linear transformations, or Bogoliubov transformations \cite{Birrell:Davies:1984}, which read
\begin{align}\label{bogoliubov:transformations}
\tilde{\mathbb{X}}=\hat{U}^\dag\,\mathbb{X}\,\hat{U}=\boldsymbol{S}\,\mathbb{X}=
\begin{pmatrix}
\boldsymbol{\alpha} & \boldsymbol{\beta} \\
\boldsymbol{\beta}^* & \boldsymbol{\alpha}^* 
\end{pmatrix}
\mathbb{X},
\end{align}
where we have introduced the vectors of operators 
\begin{align}
\mathbb{X}=&(\hat a_{\boldsymbol{1}},\dots,\hat  a_{\boldsymbol{n}},\dots;\hat a^{\dag}_{\boldsymbol{1}},\dots, \hat a^{\dag}_{\boldsymbol{n}},\dots)^{\text{T}}\nonumber\\
\tilde{\mathbb{X}}=&(\hat b_{\boldsymbol{1}},\dots, \hat b_{\boldsymbol{n}},\dots; \hat b^{\dag}_{\boldsymbol{1}},\dots, \hat b^\dag_{\boldsymbol{n}},\dots)^{\text{T}},
\end{align} 
$\hat{U}^\dag$ implements the Bogoliubov transformation and Tp stands for transposition. The matrices $\boldsymbol{\alpha}$ and $\boldsymbol{\beta}$ in the Bogoliubov transformation \eqref{bogoliubov:transformations} collect the Bogoliubov coefficients $\{\alpha_{\boldsymbol{n}\boldsymbol{m}}\}$ and $\{\beta_{\boldsymbol{n}\boldsymbol{m}}\}$ defined by $\alpha_{\boldsymbol{n}\boldsymbol{m}}:=(\phi_{\boldsymbol{n}},\tilde{\phi}_{\boldsymbol{m}})$ and $\beta_{\boldsymbol{n}\boldsymbol{m}}:=(\phi_{\boldsymbol{n}},\tilde{\phi}^*_{\boldsymbol{m}})$, where $(\cdot,\cdot)$ is the conserved inner product \cite{Birrell:Davies:1984}. These matrices satisfy the following identities $\boldsymbol{\alpha}\,\boldsymbol{\alpha}^{\dag}-\boldsymbol{\beta}\,\boldsymbol{\beta}^{\dag}=\mathds{1}$ and $\boldsymbol{\alpha}\,\boldsymbol{\beta}^{\text{T}}-\boldsymbol{\beta}\,\boldsymbol{\alpha}^{\text{T}}=0$, known as Bogoliubov identities \cite{Birrell:Davies:1984}.

\subsection{Continuous Variables and Covariance Matrix formalism}
Among all possible states $\hat \rho$ in the Hilbert space of a bosonic field, we choose to restrict ourselves to the class of Gaussian states, defined as those states with a Gaussian characteristic Wigner distribution \cite{Adesso:Ragy:2014}. These states are easily produced in the laboratory by employing standard quantum optical technologies and are at the core of many applications of quantum optics and quantum information \cite{Wang:Hiroshima:2007}. One of their main properties is that they preserve their Gaussian character under linear transformations, such as the Bogoliubov transformations \eqref{bogoliubov:transformations}. Our choice will allow us to employ the powerful tools developed in the field of continuous variables and known as Covariance Matrix formalism \cite{Adesso:Ragy:2014}. 

The first step is to realise that an infinite dimensional Gaussian state $\hat \rho$ of a bosonic field can be characterised by a finite amount of degrees of freedom i.e., its first and second moments. We conveniently choose to collect the operators of the field in the vector $\mathbb{X}$ which allows us to write compact expression for the first and second moments of the field. The canonical commutation relations take the form $[\mathbb{X}_n,\mathbb{X}^{\dag}_m]=i\,\Omega_{nm}$, where we introduce the symplectic form $\boldsymbol{\Omega}$ which, with our choice of ordering of the operators, reads   $\boldsymbol{\Omega}=\text{diag}(-i,-i,\ldots;i,i,\ldots)$. The first moments are the elements of the vector of expectation values $\langle\mathbb{X}\rangle$, while the second moments $\sigma_{nm}$ are defined by $\sigma_{nm}:=\langle\{\mathbb{X}_n,\mathbb{X}_m^{\dag}\}\rangle-2\,\langle\mathbb{X}_n\rangle\,\langle\mathbb{X}_m^{\dag}\rangle$. Here, $\{\cdot,\cdot\}$ is the anticommutator and expectation values are intended with respect to the state $\hat \rho$. The second moments can be conveniently collected in the Covariance Matrix $\boldsymbol{\sigma}$, which must obey the condition $\boldsymbol{\sigma}+i\,\boldsymbol{\Omega}\geq0$ in order for the covariance matrix to represent a physical state \cite{Adesso:Ragy:2014}.

The initial state $\hat \rho_i$ of the system evolves into the final state $\hat \rho_f$ through the standard Heisenberg equation $\hat \rho_f=\hat U\,\hat \rho_i\,\hat U^{\dag}$, where the unitary operator $\hat U$ encodes the transformation that one is considering. If the transformation is linear (i.e., quadratic in the creation and annihilation operators, or equivalently in the quadrature operators), it is possible to represent the operator $\hat U$ by a symplectic matrix $\boldsymbol{S}$ which preserves the symplectic form $\boldsymbol{\Omega}$, i.e., $\boldsymbol{S}^{\dag}\,\boldsymbol{\Omega}\,\boldsymbol{S}=\boldsymbol{S}\,\boldsymbol{\Omega}\,\boldsymbol{S}^{\dag}=\boldsymbol{\Omega}$. The Heisenberg equation then reads $\boldsymbol{\sigma}_f=\boldsymbol{S}\,\boldsymbol{\sigma}_i\,\boldsymbol{S}^{\dag}$, where $\boldsymbol{\sigma}_i$ and $\boldsymbol{\sigma}_f$ represent the initial and final state respectively. Linear transformations are well known operations in quantum optics; for example one has beam splitting and single- and two-mode squeezing \cite{Mandel:Wolf:1995}. Bogoliubov transformations are linear transformations, which implement the effects of spacetime dynamics or moving boundary conditions \cite{Birrell:Davies:1984}. An arbitrary Bogoliubov transformation can be represented by a symplectic matrix $\boldsymbol{S}$ as well. It can be easily shown that the symplectic matrix $\boldsymbol{S}$ that represents a Bogoliubov transformation is exactly the matrix that appears in equation \eqref{bogoliubov:transformations}. Bogoliubov transformations are represented by symplectic matrices $\boldsymbol{S}$ that preserve the symplectic form. It is easy to see that this property is equivalent to the well known Bogoliubov identities $\boldsymbol{\alpha}\,\boldsymbol{\alpha}^{\dag}-\boldsymbol{\beta}\,\boldsymbol{\beta}^{\dag}=\mathds{1}$ and $\boldsymbol{\alpha}\,\boldsymbol{\beta}^{\text{T}}-\boldsymbol{\beta}\,\boldsymbol{\alpha}^{\text{T}}=0$ listed before, see also \cite{Birrell:Davies:1984}. Altogether, this introduction to Covariance Matrix formalism shows why it is convenient to restrict our analysis to Gaussian states. 

\subsection{Perturbative approach}\label{subsection:peturbative approach}
It often occurs in physical scenarios that results can be obtained only within suitable perturbative approaches. This is, typically, also a symptom of the fact that physical control over a particular process exists \textit{only} within these perturbative regimes, such as weak couplings. Outside these regimes, dynamics become uncontrollable. Important examples are, for example, the dynamical Casmir effect \cite{Dodonov:2010}, optomechanical systems \cite{Aspelmeyer:Kippenberg:2014}, parametric down conversion \cite{Walborn:Monken:2010} and gravitational waves \cite{Abbott:Abbott:2016}.

In this work we will find general expressions for the efficiency of cavities as thermodynamical machines in terms of the coefficients of the Bogoliubov transformations, which encode the time evolution of the system and all spacetimes parameters. However it is not possible in general to find an analytical expression of the Bogoliubov coefficients of interest, especially when time dependence is non trivial. Therefore, we will specialise to cavities whose boundaries (i.e., boundary conditions) move in flat spacetime. This includes the  effects of a gravitational wave on the phononic field of a BEC \cite{Sabin:Bruschi:2014}. A well known application of this approximation is the theory behind the laser-interferometer gravitational wave detectors, such as LIGO \cite{Abbott:Abbott:2016}. The theory of cavities that move with small but arbitrarily changing proper accelerations, or that are subject to transformations which encode a small physical parameter to be estimated, has been developed in the past few years, with applications to superconducting circuits \cite{Wilson:Johansson:2011} and detection of gravitational waves \cite{Sabin:Bruschi:2014}.

Every quantity of interest in this work will be expanded as a function of a small dimensionless parameter $h\ll1$. Below we provide a few examples
\begin{align}\label{perturbative:regime}
\boldsymbol{S}=&\,\boldsymbol{S}^{(0)}+\boldsymbol{S}^{(1)}\,h+\boldsymbol{S}^{(2)}\,h^2+\mathcal{O}(h^3)\nonumber\\
\alpha_{\boldsymbol{n}\boldsymbol{m}}=&\,\alpha^{(0)}_{\boldsymbol{n}\boldsymbol{m}}+\alpha^{(1)}_{\boldsymbol{n}\boldsymbol{m}}\,h+\alpha^{(2)}_{\boldsymbol{n}\boldsymbol{m}}\,h^2+\mathcal{O}(h^3)\nonumber\\
\beta_{\boldsymbol{n}\boldsymbol{m}}=&\,\beta^{(1)}_{\boldsymbol{n}\boldsymbol{m}}\,h+\beta^{(2)}_{\boldsymbol{n}\boldsymbol{m}}\,h^2+\mathcal{O}(h^3)\nonumber\\
\boldsymbol{\sigma}=&\,\boldsymbol{\sigma}^{(0)}+\boldsymbol{\sigma}^{(1)}\,h+\boldsymbol{\sigma}^{(2)}\,h^2+\mathcal{O}(h^3),
\end{align}
and analogously for every other quantity that is a function of the parameter $h$, such as suitable measures of entanglement \cite{Bruschi:Dragan:2013}.
The list of properties satisfied by the Bogoliubov coefficients for specific applications to moving cavities can be found in \eqref{perturbative:contributions:to:bogoliubov:identities}, see also \cite{Bruschi:Dragan:2013}.

In these perturbative cavity scenarios it is possible to show that the frequency $\omega_{\boldsymbol{n}}$ of the modes takes the simple expected expression $\omega_{\boldsymbol{n}}\equiv\omega_{n,m,p}=\sqrt{(\frac{\pi\,n\,c}{L})^2+(\frac{\pi\,m}{L_y})^2+(\frac{\pi\,p}{L_z})^2}+\mathcal{O}(h^3)$. Finally, the Heisenberg relation between the initial state $\boldsymbol{\sigma}(0)$ and the final state $\boldsymbol{\sigma}(h)$ in the covariance matrix formalism will read $\boldsymbol{\sigma}(h)=\boldsymbol{S}(h)\,\boldsymbol{\sigma}(0)\,\boldsymbol{S}^{\dag}(h)$

\section{Energetics and performance of relativistic quantum machines}\label{performance}
Cavity dynamics (or dynamics of boundary conditions) excite, in general, all modes of a quantum field confined within the cavity, a phenomenon which is generally known as dynamical Casimir effect \cite{Dodonov:2010}. However, the average change in population of each mode greatly depends on the energy that is transferred into the modes. In particular, in the case of periodic modulation/motion, one expects the existence of a ``resonant'' behavior \cite{Bruschi:Dragan:2013}. Furthermore, it is in general possible to experimentally access only a (limited) part of the cavity spectrum, for example the lowest two modes, with higher energy modes leaking into the environment and being irreversibly lost. This inability to access the whole spectrum but only a few modes occurs in an many physical systems, such as superconducting circuits \cite{SandboChang:Simoen:2017}. In this work we assume that the cavity is ideal, that is, its modes aren't allowed to leak into the environment. Attempts to explicitly incorporate the environment into our dynamics would require us to consider appropriate master equations \cite{kosloff2013quantum}. We leave it to future work to employ open quantum systems techniques to address this important aspect.

\subsection{Accessible modes: choice of the system and the environment}
As only a few modes of the system can be accessed we choose to divide the mode spectrum of the cavity into two parts: the System S, which is the part of the spectrum that can be experimentally accessed and manipulated, and the Environment E, which is composed of all the other cavity modes, whether inaccessible to experimental probes or those that leak outside of the cavity faster than any detection time. We also define the Cavity C as the set of all modes within the cavity (in this case $k_x\geq1$,$k_y\geq1$ and $k_z\geq1$), i.e., the union of S and E.

For the sake of simplicity, in the following we focus on cavities that are effectively $1$-dimensional. We notice that the transverse dimensions $(y,z)$ can be reduced to an effective mass $M_{eff}$, which contributes to the one dimensional frequency in a standard way \cite{Bruschi:Fuentes:2012}. We can further assume that the cavity is small in the transverse dimensions as compared to the one of interest. Therefore, initially unpopulated transverse modes remain unpopulated due to the higher energies necessary to excite them. We can replace then all indices of the form $\boldsymbol{n}=(n,m,p)$ with $n$, and set $L_x=L$. For example, the mode frequency is now $\omega_n=\frac{\pi\,n\,c}{L}+\mathcal{O}(h^3)$, see \cite{Bruschi:Fuentes:2012}.  

\subsection{Initial state}
We assume that the cavity is \textit{not} in contact with a thermal bath B of temperature $T$, but that the cavity modes can have been prepared initially in a thermal state of temperature $T$. In this sense, the system is isolated and cannot leak energy or excitations. This, again, is an idealization of realistic implementations but it allows us to focus our study on the energy transfer between the degrees of freedom of interest and the excitations of the field. For this reason, standard unitary evolution of the whole cavity C is assumed.

\subsection{Energy and entropy in a cavity}

The average energy $E_k$ of a mode $k$ is $E_k:=\hbar\,\omega_k\,N_k$, where $N_k=\frac{\sigma_{kk}-1}{2}$ is the number expectation value of mode $k$ expressed in terms of elements of the covariance matrix when the first moments of the state are zero. The total energy $E$ contained in the field at any time is simply given by $E=\sum_{k\in C}E_k$. The total energy $\Delta E_C$ \textit{absorbed} by the whole field in the cavity after a Bogoliubov transformation is defined as $\Delta E_C:=E_f-E_i$, where $E_f$ and $E_i$ are the average energy content of the field before and after the Bogoliubov transformation respectively. In the covariance matrix formalism, the total absorbed energy $\Delta E_C$ takes the expression
\begin{align}\label{energy:expression:covariance:matrix}
\Delta E_C=\hbar\sum_{k\in C}\omega_k\,\frac{\sigma_{f,kk}-\sigma_{i,kk}}{2}.
\end{align}
Analogously, the change of energy of the system $\Delta E_\textrm{S}$ and the environment $\Delta E_E$ are readily found by replacing $C$ with $S$ and $E$ and the summation is performed over the modes in the relevant set. Clearly, $\Delta E_C=\Delta E_\textrm{S}+\Delta E_E$.

We now proceed to compute the relevant changes in entropy.  As we have already mentioned, the total change of entropy vanishes because Bogoliubov transformations are unitary transformations on the whole cavity C. However, the change in Von Neumann entropy $\Delta S_\textrm{S}$ of the systems S can be found by employing known expressions \cite{Adesso:Ragy:2014}. It reads
\begin{align}\label{von:neumann:entropy}
\Delta S_\textrm{S}=k_B\,\sum_{k\in S}\left[f_+(\nu_k)-f_-(\nu_k)\right],
\end{align}
where $\nu_k$ are the symplectic eigenvalues of the state (i.e., the eigenvalues of $i\,\boldsymbol{\Omega}\,\boldsymbol{\sigma}$, see \cite{Adesso:Ragy:2014}) and $f_{\pm}(x):=\frac{x\pm1}{2}\,\ln(\frac{x\pm1}{2})$.  Note that these unitary Bogoliubov transformations are not restricted to any one class of thermal unitary operations, such as cyclic unitaries \cite{allahverdyan2004maximal} or the free unitaries in the resource theory of thermodynamics \cite{brandao2013resource}, but instead are kept unrestricted. This choice allows us to study the most general situation in which the classical gravitational degrees of freedom may indeed change the Hamiltonian of the system in order to impart energy into the quantum field. 

\subsection{Efficiency}
Equipped with all the tools described before we can now investigate the performance of confined localised relativistic quantum fields as thermal machines.
We start by defining the relevant figure of merit that allows us to quantify such performance. Along these lines we introduce the \textit{efficiency} $\eta$, defined as the following  fraction, 
\begin{align}
\eta:=\frac{\Delta E_\textrm{S}}{\Delta E_\textrm{C}},
\end{align}
where $\Delta E_\textrm{S}$ and $\Delta E_\textrm{C}$ are the total change of energy in the system S and cavity C respectively. Using the first and second laws of thermodynamics we know that the total extractable work $W$ from a system undergoing a  unitary evolution is upper bounded by the change in energy of the total cavity, 
\begin{align}
	W\leq\Delta E_\textrm{S}=\eta\Delta E_\textrm{C},
\end{align}
demonstrating the importance of quantifying the \textit{efficiency} of said system. Noting that $\Delta E_\textrm{S}+\Delta E_\textrm{E}=\Delta E_\textrm{C}$, we obtain
\begin{align}\label{efficiency}
\frac{1}{\eta}=1+\frac{\Delta E_\textrm{E}}{\Delta E_\textrm{S}}.
\end{align}
We keep the inverse of the efficiency for convenience of presentation.

The efficiency \eqref{efficiency} is therefore directly related with the increase or decrease of entropy of the system's state, as expected.
More explicitly, equation \eqref{efficiency} can be written as
\begin{align}\label{efficiency:two}
\frac{1}{\eta}=1+\frac{\sum_{k\in E}k\,(\Delta N^{(1)}_k+\Delta N^{(2)}_k\,h)}{\sum_{k\in S}k\,(\Delta N^{(1)}_k+\Delta N^{(2)}_k\,h)}.
\end{align}
We proceed to compute explicitly all terms in \eqref{efficiency:two} to lowest order in the perturbative parameter $h$. In particular, we have already anticipated that $\Delta N^{(0)}_k=0$.

\subsection{Efficiency behavior}
We are in the position to anticipate that there are two different possible behaviors of the efficiency \eqref{efficiency:two}, which we list below:
\begin{itemize}
	\item[i)] The contributions $\Delta N^{(1)}_k$ vanish both in the numerators and denominator, which would give us an efficiency of the form
	\begin{align}\label{perturbative:efficiency:case:i}
		\frac{1}{\eta}&=1-\Gamma(t)+\mathcal{O}(h),
	\end{align}
where $\Gamma(t)$ is a time dependent function independent of $h$.
	\item[ii)] The term $\Delta N^{(1)}_k$ vanishes in the numerator but not in the denominator. Therefore, expression \eqref{efficiency:two} reads
\begin{align}\label{perturbative:efficiency:case:ii}
\frac{1}{\eta}&=1-\Gamma^{\prime}(t)\,h+\mathcal{O}(h^2),
\end{align}
where $\Gamma^{\prime}(t)$ is a time dependent function independent of $h$.
\end{itemize}
We notice, surprisingly, that the two possible scenarios are markedly different. Note that it other behaviours are not possible due to the fact that the energy is a continuous and finite function of the parameters of the system in this perturbative regime.
 
In section \ref{section:role:initial:state} we will proceed to  show how can the different scenarios described above arise due to the choice of the initial Gaussian state.

\subsection{Thermodynamic fluctuations in the cavity}
One of the central results of stochastic thermodynamics was the discovery of fluctuation relations governing the behaviour of thermodynamic variables giving corrections to the classical laws of thermodynamics \cite{jarzynski1997nonequilibrium, crooks1999entropy}. Given the fact that stochastic thermodynamics is now being seen as the stepping stone from a classical understanding of thermodynamics to a quantum one, several recent studies have tried to investigate the role of these fluctuations in a quantum setting \cite{aaberg2018fully,holmes2019coherent,mingo2019decomposable,alhambra2019entanglement,alhambra2016fluctuating,morris2018quantum}. This raises an important question regarding our system \textit{``Do work fluctuations play a role in relativistic cavity dynamics?"} And, if this is the case, \textit{``what is the physical situation in which they emerge?"} This is the one of the questions that we will address in this work.

The central object to fluctuation relations is the \emph{forward work probability distribution} $P(W)$, which describes the probability of the system to evolve from an initial state $\hat{\rho}_{\textrm{i}}$ to a final state $\hat{\rho}(\lambda)$ by absorbing the amount of work  $W$ in the process \cite{Goold:Huber:2016}. Given a Hamiltonian $\hat{H}$, an initial state $\hat{\rho}_{\textrm{i}}$ and a final state $\hat{\rho}(\lambda)$, the work probability distribution is defined as
\begin{align}\label{f_prob} 
P_F(W)=\sum_{mn}\,p_{m|n}\,p_n\,\delta\big{(}W-[E^{f}_{m}-E^{i}_{n}]\big{)}.
\end{align}    
Here $p_n:=\text{Tr}(\hat{\rho}_i\,|n\rangle\langle n|)$ is the probability of finding the initial state $\hat{\rho}_{\textrm{i}}$ of the system in an eigenstate $|n_p\rangle$ of the Hamiltonian $\hat{H}$. The values $E^{i}_{n}=n\,\omega(0)$ and $E^{f}_{m}=n\,\omega(\lambda)$ are the $n$th and $m$th eigenvalues of the initial and final Hamiltonian respectively. The probability $p_{m|n}$ is the conditional probability that the unitary evolution of the initial eigenstate $|n\rangle_i$ results in the final state ${}_f\langle m|$. Here, $\delta(x)$ stands for the Dirac delta.

The physical or operational interpretation of the forward probability work distribution is defined via the following protocol: 
\begin{itemize}
	\item[i)] Perform a projective measurement $|n\rangle_i\langle n|$ on the initial state $\hat{\rho}_{\textrm{i}}$. Obtain the value $p_n:=|{}_i\langle n|\,\rho_i\,|n\rangle_i|^2$.
	\item[ii)] Transform (evolve) the initial state $\hat{\rho}_{\textrm{i}}$ through a unitary transformation $\hat{U}(\lambda)$ and obtain the final state $\hat{\rho}(\lambda)=\hat{U}(\lambda)\,\hat{\rho}_{\textrm{i}}\,\hat{U}^\dag(\lambda)$.
	\item[iii)] Perform a projective measurement $|m\rangle_f\langle m|$ on the final state $\hat{\rho}(\lambda)$. Obtain the value $p_{m|n}:=|{}_f\langle m|\,\hat{U}(\lambda)\,|n\rangle_i|^2$.
\end{itemize}
The forward probability distribution $P_F(W)$ has a clear thermodynamical interpretation \cite{esposito2009nonequilibrium}. It defines the average work performed on the system in the forward process $\hat{\rho}_{\textrm{i}}\rightarrow\hat{\rho}(\lambda)$, since 
\begin{align*}
\langle W\rangle:=&\int dW\,W\,P_F(W)\nonumber\\
=&\sum_{mn}\,p_{m|n}\,p_n\,(E^{f}_{m}-E^{i}_{n})
\end{align*} 
and the \emph{total} change of entropy vanishes (i.e., the transformation is unitary).  

This relation is a stochastic relation, which takes into account that, quantum mechanically, there is no well defined, or a-priori special, transition that occurs as a result of ``arbitrary'' unitary processes but rather the evolution can cause the system to take any one of the discrete values of work with some probabilistic weighting. It is stressed that the results presented here are not restricted to any one particular choice of initial state but instead are completely general in this regard. In order to recover the celebrated Jarzynski and Tasaki-Crooks fluctuation relations  \cite{jarzynski1997nonequilibrium, crooks1999entropy}, one need only to specify that the initial state is in thermal equilibrium with its environment. 

Analytically dealing with the Dirac $\delta$ distribution in \eqref{f_prob}  is hard. Therefore, it is convenient to introduce the \emph{characteristic function} $\chi(u):=\int dW\,\exp[-i\,u\,W]\,P_F(W)$, i.e., the Fourier transform of \eqref{f_prob}. By employing \eqref{f_prob} , this reads 
\begin{align}\label{characteristic:function}
\chi(u):=\sum_{m n}\,p_{m|n}\,p_n\,e^{-i\,(E^{f}_{m}-E^{i}_{n})\,u}.
\end{align}  
The characteristic function contains useful information about the system and the process considered. If it is possible to obtain the analytical expression of $\chi(u)$, it is then immediately possible to obtain any moment of the stochastic distribution of work. In particular, the $n$th moment of the work probability distribution is
\begin{align}\label{moment}
G[\psi^n]=i^{n}\,\Big{[}\frac{d^n}{du^n}\,\chi_\psi (u)\Big{]}_{u=0}.
\end{align}
The first moment ($n=1$) is the mean energy change in the process, while the second moment ($n=2$) gives the variance of the distribution. The third moment ($n=3$) is known as the skewness, describing the asymmetry of the distribution. The aim of this work is to derive the moments of the work distribution in terms of the relativistic cavity dynamics. 
\section{Efficiency: the role of the initial state}\label{section:role:initial:state}

Here we proceed to give the results for the different initial states. We choose four initial Gaussian states of particular interest for their paradigmatic importance in quantum optics for the system S: a thermal state, a thermal beam-splitted state of two modes, a two-mode thermal state of individually squeezed modes and finally a two mode thermal squeezed state. We emphasize that the remaining modes of the field, i.e., those belonging to the environment E, are assumed to be in a thermal state.

\subsection{Efficiency: no diagonal Bogoliubov coefficients}

We start by assuming that, to first order, there are no diagonal Bogoliubov coefficients, that is, $\alpha_{k\tilde{k}}^{(1)}=\beta_{k\tilde{k}}^{(1)}=0$. This is justified in a wide class of scenarios from previous work \cite{Bruschi:Louko:2013}.

\subsubsection{Efficiency: initial thermal state}
Here the initial state of the field is a thermal state $\boldsymbol{\sigma}_{th}$. In the Covariance Matrix formalism it takes the expression $\boldsymbol{\sigma}_{th}=\text{diag}(\nu_1,\nu_2,\ldots;\nu_1,\nu_2,\ldots)$, where $\nu_k=\coth(\frac{\hbar\,\omega_k}{k_B\,T})$ are the symplectic eigenvalues of the state and $\boldsymbol{\sigma}_{th}$ is equal to its Williamson form $\boldsymbol{\sigma}_{\oplus}$ i.e., $\boldsymbol{\sigma}_{th}=\boldsymbol{\sigma}_{\oplus}$ see \cite{Adesso:Ragy:2014}. 

In \ref{appendix:three} we derive the efficiency for the initial thermal state $\boldsymbol{\sigma}_{th}$, which reads
\begin{align}\label{perturbative:efficiency:initial:thermal:states}
\frac{1}{\eta}=&1+\frac{\sum_{k_\text{E}\in E}\sum_{p\in\text{C}}\,k_\text{E}\,\mathcal{Z}_{k_\text{E}p}(t)}{\sum_{k\in S}\sum_{p\in\text{C}}\,k\,\mathcal{Z}_{kp}(t)},
\end{align}
where we have introduced the function 
\begin{align*}
\mathcal{Z}_{nm}(t):=|\beta^{(1)}_{nm}|^2\,(\nu_n+\nu_m)+|\alpha^{(1)}_{nm}|^2\,(\nu_n-\nu_m).
\end{align*}
The efficiency \eqref{perturbative:efficiency:initial:thermal:states} takes the form anticipated in scenario i). Corrections to this expression come to first order in $h$.

\subsubsection{Efficiency: thermal beam-splitted states}
Let us now turn to a system S of two modes $k$ and $\tilde{k}$. For simplicity, we assume that each mode $k_\text{E}$ of the environment E is in a thermal state. Instead, the initial state $\boldsymbol{\sigma}_{i,\text{BS}}(\theta)$ of the system S is defined by the elements $U_{\text{i},kk}$, $U_{\text{i},k_\text{E}k_\text{E}}$ and $U_{\text{i},kk_\text{E}}=U_{\text{i},k_\text{E}k}$, while $V_{\text{i},km}\equiv0$. We compute the efficiency in \ref{appendix:three} and find
\begin{align}\label{beam:splitted:state:perturbative:efficiency}
\frac{1}{\eta}=&1+\frac{\sum_{k_\text{E}\in E}\sum_{p\in\text{C}}\,k_\text{E}\,\mathcal{Z}_{k_\text{E}p}(t)}{(k-\tilde{k})(\nu_k+\nu_{\tilde{k}})\,\sin(2\,\theta)\Re\left(\alpha_{kk}^{(0)}\,\alpha_{k\tilde{k}}^{(1)*}\right)}\,h.
\end{align}
Notice that $\theta\gg h$ in our calculations above. For the cases where $\theta\sim h$ or $\theta\ll h$ we have to proceed differently -- the perturbative expansion has to be performed around $\theta=0$ rather than around $h=0$.

The efficiency \eqref{beam:splitted:state:perturbative:efficiency} takes the form anticipated in scenario ii). Corrections to this expression come to first order in $h^2$.

\subsubsection{Efficiency: single-mode squeezed states}
Let us now turn to a single mode squeezed state $\boldsymbol{\sigma}_{i,\text{SMS}}(s)$ of two initially thermal modes $k$ and $\tilde{k}$ that compose our system S. For simplicity, we assume that both modes of the system S are single-mode squeezed with squeezing parameter $s$. The initial state $\boldsymbol{\sigma}_{i}$ now has nonzero elements $U_{\text{i},kk}=\nu_k\,\cosh (2\,s)$, $U_{\text{i},\tilde{k}\tilde{k}}=\nu_{\tilde{k}}\,\cosh (2\,s)$, and $V_{\text{i},kk}=\nu_k\,\sinh (2\,s)$ and $V_{\text{i},\tilde{k}\tilde{k}}=\nu_{\tilde{k}}\,\sinh (2\,s)$, while all others vanish. We compute the efficiency in \ref{appendix:three} and find
\begin{align}\label{single:mode:squeezed:state:perturbative:efficiency}
\frac{1}{\eta}=&1+\frac{P_\text{E}(t)}{P_\text{S}(t)},
\end{align}
where the functions $P_\text{E}(t)$ and $P_\text{S}(t)$ are defined in \eqref{p:functions:appendix} for simplicity.
As expected, when $s=0$ we recover the initial thermal state efficiency \eqref{perturbative:efficiency:initial:thermal:states}.

The efficiency \eqref{single:mode:squeezed:state:perturbative:efficiency} takes the form anticipated in scenario i). Corrections to this expression come to first order in $h$.

\subsubsection{Efficiency: two-mode squeezed states}
We finally turn our attention to initial states that are two-mode squeezed. For simplicity of presentation of our techniques, we assume again that the system S is composed of two modes $k$ and $\tilde{k}$ only and that these modes are initially in a two mode squeezed state $\boldsymbol{\sigma}_{i}(r)$ with squeezing parameter $r$. Generalization to multimode squeezing is straightforward and does not yield any conceptual novelty.

In \ref{appendix:three} we compute the efficiency for this case which reads
\begin{align}\label{two:mode:squeezed:state:perturbative:efficiency}
\frac{1}{\eta}=&1+\frac{\sum_{k_E\in E}\,k_\text{E}\,U^{(2)}_{f,k_\text{E}k_\text{E}}}{(k+\tilde{k}) (\nu_k+\nu_{\tilde{k}}) \Re\left(\alpha_{kk}^{(0)}\,\beta_{k\tilde{k}}^{(1)*}\right) \sinh(2\,r)} h,
\end{align}
where the function $U^{(2)}_{f,k_\text{E}k_\text{E}}$ is defined in \eqref{U2:function:appendix}.
Note the dependence of \eqref{two:mode:squeezed:state:perturbative:efficiency} on the squeezing. 

The efficiency \eqref{two:mode:squeezed:state:perturbative:efficiency} takes the form anticipated in scenario ii). Corrections to this expression come to order $h^2$.

In the following, we proceed to discuss the results obtained so far in the limit of low initial temperatures, which allows us to simplify our results o some degree.

\subsubsection{Efficiency: low temperatures}
Let us comment on the low temperature regime, that is, when $\nu_n\sim1$ for all $n$. This also means that $k_\text{B}\,T\ll\hbar\,\omega_1$. This is the case, for example, for state of the art Bose Einstein condensation \cite{Dalfovo:Giorgini:1999}.

Below we list the four efficiencies that we have computed above, evaluated in this regime. We find
\begin{align}
\frac{1}{\eta}\sim&1+\frac{\sum_{k_\text{E}\in E}\sum_{p\in\text{C}}\,k_\text{E}\,|\beta^{(1)}_{k_\text{E}p}|^2}{\sum_{k\in S}\sum_{p\in\text{C}}\,k\,|\beta^{(1)}_{kp}|^2}\nonumber\\
\frac{1}{\eta}\sim&1+\frac{\sum_{k_\text{E}\in E}\sum_{p\in\text{C}}\,k_\text{E}\,|\beta^{(1)}_{k_\text{E}p}|^2}{(k-\tilde{k})\,\sin(2\,\theta)\,\Re\left(\alpha_{kk}^{(0)}\,\alpha_{k\tilde{k}}^{(1)*}\right)}\,h\nonumber\\
\frac{1}{\eta}\sim&\left.1+\frac{P_\text{E}(t)}{P_\text{S}(t)}\right|_{\nu_n=1\,\,\forall n}\nonumber\\
\frac{1}{\eta}\sim&1+\frac{\sum_{k_E\in E}\,k_\text{E}\,U^{(2)}_{f,k_\text{E}k_\text{E}}}{(k+\tilde{k}) \Re\left(\alpha_{kk}^{(0)}\,\beta_{k\tilde{k}}^{(1)*}\right) \sinh(2\,r)} h&
\end{align}
where here we have defined the simplified functions
\begin{align}
U^{(2)}_{f,k_\text{E}k_\text{E}}:=&\left[|\beta_{k_\text{E}k}^{(1)}|^2+|\beta_{k_\text{E}\tilde{k}}^{(1)}|^2\right]\,\cosh^2r\nonumber\\
&+\left[|\alpha_{k_\text{E}k}^{(1)}|^2+|\alpha_{k_\text{E}\tilde{k}}^{(1)}|^2\right]\,\sinh^2r\nonumber\\
&+\Re\left(\alpha_{k_\text{E}k}^{(1)}\,\beta_{k_\text{E}\tilde{k}}^{(1)*}\right)\,\sinh(2\,r)\nonumber\\
&+\Re\left(\alpha_{k_\text{E}\tilde{k}}^{(1)}\,\beta_{k_\text{E}k}^{(1)*}\right)\,\sinh(2\,r)\nonumber\\
&+\sum_{p\in E}|\beta_{k_\text{E}p}^{(1)}|^2.
\end{align}
The simplified equations above give us more insight into the efficiency when temperatures are lower than the energy necessary to create an excitation.

\subsection{Efficiency: nonzero diagonal Bogoliubov coefficients}

We repeat our calculations assuming that, to first order, there exist diagonal Bogoliubov $\beta$ coefficients and these are nonzero. This is justified in another wide class of scenarios from previous work \cite{Barbado:Baez-Camargo:2018}, namely cavities with ``free falling'' boundaries under motion or dynamics of background spacetimes.

We note that, in this case, only contributions that are of first order in the perturbative parameter $h$ have to be retained. We consider three initial states: the thermal state, the thermal beam-splitted state and the single mode squeezed state of one mode. The crucial term to consider is
\begin{align}
U_{S,nn}^{(1)}=&2\,\sum_{p\in\text{C}}\,\Re\left(\alpha^{(0)*}_{nn}\,\alpha^{(1)}_{np}\,U_{\text{i},pn}\right)\nonumber\\
&+2\,\sum_{p\in\text{C}}\,\Re\left(\alpha^{(0)}_{nn}\,\beta^{(1)*}_{np}\,V_{\text{i},np}\right).
\end{align}
Below we consider the explicit case of the desired initial states.

\subsubsection{Initial thermal state and beam-splitted state}
In both of these cases, the situation is similar to the previous one, since it is simple to show that the first order terms $U_{S,nn}^{(1)}$ vanish.
One needs to proceed to second order, and it is also easy to check that this allows us to recover the previous results. In this sense, no difference occurs between the two types of Bogoliubov coefficients.

\subsubsection{Initial single mode squeezed thermal state}
The case is different for a single mode squeezed thermal state of mode $k$, that has the form
\begin{align}
\boldsymbol{\sigma}_{i,\text{SMS}}(s)=
\nu_k\,\begin{pmatrix}
\cosh (2\,s) & \sinh (2\,s)\\
\sinh (2\,s) & \cosh (2\,s)
\end{pmatrix}.
\end{align}
We find here that, already with one mode, there is an interesting occurrence and find
\begin{align}\label{energy:expression:covariance:matrix:appendix:squeezed}
\frac{1}{\eta}=&1+\frac{\sum_{k_\text{E}\in E}\,k_\text{E}\,U^{(2)}_{f,k_\text{E}k_\text{E}}}{\nu_k\,\Re\left(\alpha^{(0)*}_{kk}\,\beta^{(1)}_{kk}\right)\,\sinh(2\,s)}\,h,
\end{align}
where we have conveniently introduced the following expression
\begin{align}\label{U2:function}
U^{(2)}_{f,k_\text{E}k_\text{E}}:=&|\beta_{k_\text{E}k}^{(1)}|^2(\nu_k\,\cosh(2\,s)+\nu_{k_\text{E}})\nonumber\\
&+|\alpha_{k_\text{E}k}^{(1)}|^2(\nu_k\,\cosh(2\,s)-\nu_{k_\text{E}})\nonumber\\
&+\nu_k\sinh(2\,s)\,\Re\left(\alpha_{k_\text{E}k}^{(1)}\,\beta_{k_\text{E}k}^{(1)*}\right)\nonumber\\
&+\sum_{p\in E}\mathcal{Z}_{k_\text{E}p}(t)
\end{align}
for ease of presentation.

Notice the similarity between this result and the one obtained above for the two mode squeezed state.

\section{Charging a quantum battery through motion and gravity: bounds on the extractible energy}\label{charging:battery}
We have computed the efficiency of energy transfer for an isolated cavity, which we have defined as the ratio between the amount of energy transferred to the system S and the energy transferred to to whole cavity modes C. This, however, provides only an upper bound to the amount of energy that we can potentially extract from the system S through some appropriate mechanism.

We now proceed to employ an existing scheme to extract the energy stored in the system S, and we then compute the efficiency of the extraction process. This scheme comprises of two cavities R (reservoir) and H (hot), which will act as our active systems, and a harmonic oscillator B, which will act as the battery. The system S in the cavity H will be composed of one mode, labeled by the quantum number $n$. This is necessary for later discussion of Carnot bounds on performance \cite{Scully:Zubairy:2003}. If we started with more than one mode in system S such modes would, in general, have different temperatures after the cavity H has undergone a Bogoliubov transformation and they would be correlated. This can, in principle, lead to local violations of the Carnot bound and it will not allow us to obtain analytical expressions \cite{Scully:Zubairy:2003}.
The protocol that we will employ is illustrated in the box provided opposite.

We now proceed to describe each step of the protocol. 
\tcbset{colback=red!5!white,fonttitle=\bfseries}
\begin{tcolorbox}[float=htpb!,enhanced,title=A protocol to extract energy,
frame style={left color=red!75!black,right color=blue!75!black}]\label{box}
Protocol to extract energy from a cavity that is affected by motion or gravity. This protocol has been extensively studied in literature \cite{note}.

---------------------------------------------------------

i) Cavities R and H are identical, and at the same temperature. Cavity modes are initially excited.

ii) Motion or gravity affect cavity H. Its field, at the end, is left in a different state. 

\begin{minipage}[c]{\linewidth}
    \vspace*{-40pt}
        \includegraphics[width=\linewidth]{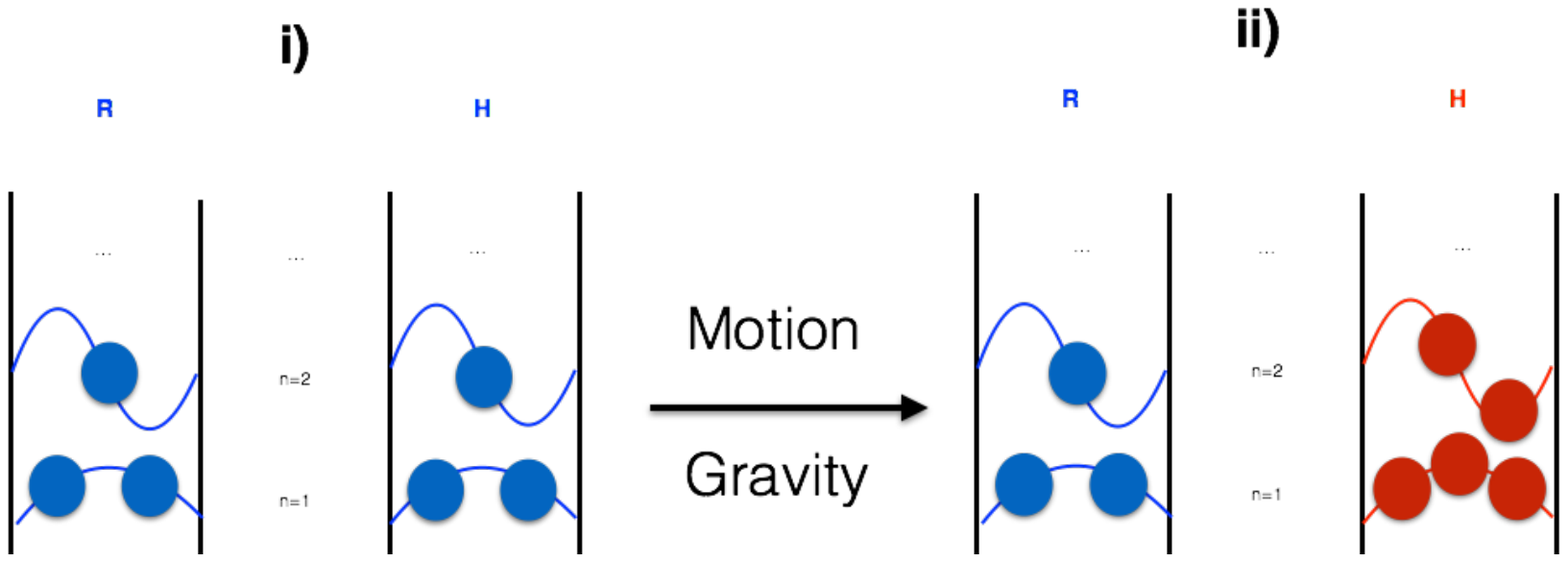}
    \end{minipage}
\vspace*{-40pt}    

---------------------------------------------------------

iii) The highlighted mode of choice of cavities R, H and battery B (as an example, mode $n=1$) interacts through the interaction Hamiltonian $H_I$.
\begin{minipage}[t]{\linewidth}
        \includegraphics[width=\linewidth, height=9.5cm]{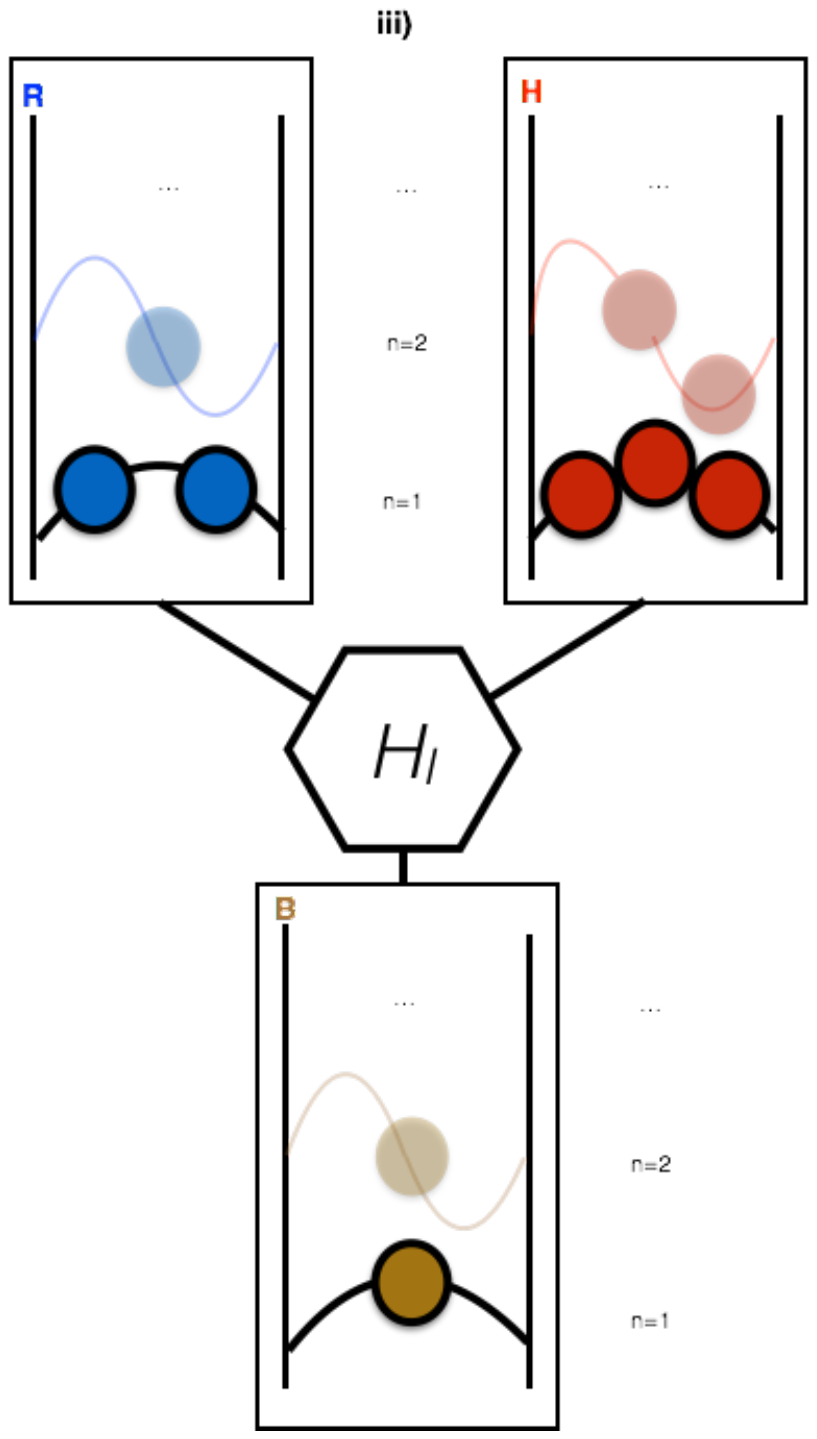}
    \end{minipage}
   Ideally, we can extract excitations from cavity $H$ and store them in the battery B.

---------------------------------------------------------

 iv) We can compute the total bound of the efficiency and obtain the final result \eqref{bound:on:energy:storage}.
\end{tcolorbox}

\begin{itemize}
	\item[i)] \textit{Initialisation}: Cavity R is left at rest as a reference cavity, or ``reservoir'' or ``cold'' system, with the whole cavity R in a thermal state $\boldsymbol{\sigma}_{R,i}=\bigoplus_n\,\nu_n\,\mathds{1}_{2\times2}$, where $\nu_n=\coth(\frac{\hbar\,\omega_n}{2\,k_\text{B}\,T})$.  Cavity H, that has the confined field initially in the same thermal state $\boldsymbol{\sigma}_{H,i}=\bigoplus_n\,\nu_n\,\mathds{1}_{2\times2}$ with temperature $T$, is affected by gravity or undergoes some motion. As a consequence, the field confined inside is now excited (the modes are populated) and therefore the single-mode reduced state $\boldsymbol{\sigma}_{H,n}(h)$ of the $n$-th mode is, in general, not thermal, i.e., $\boldsymbol{\sigma}_{H,n}(h)\neq\nu_n\,\mathds{1}_{2\times2}$. 
	\item[ii)] \textit{Available states after evolution}: The reservoir R remains in the initial thermal state $\boldsymbol{\sigma}_{R,i}$. The final one-mode reduced states $\boldsymbol{\sigma}_{H,n}(h)$ of the ``hot'' cavity, however, are all in a \textit{different} thermal state (up to local transformations). This is a direct consequence of the initial state being Gaussian and the Bogoliubov transformation being Gaussian (i.e., linear) transformations \cite{Adesso:Ragy:2014}. It is easy to show that the final one-mode reduced state $\boldsymbol{\sigma}_{H,n}(h)$ of mode $n$ has the form
	\begin{align}\label{final:single:mode:reduced:state}
	\boldsymbol{\sigma}_{H,n}(h)\sim\,
	\begin{pmatrix}
	\nu_n+2\,A_n\,h^2 & 2\,B_n\,h^2\\
	2\,B^*_n\,h^2 & \nu_n+2\,A_n\,h^2
	\end{pmatrix},
	\end{align}
	where we have introduced for the sake of convenience 
{\small
\begin{align}A_n:=&\frac{1}{2}\,\sum_{m\in C}\left[(\nu_n+\nu_m)|\beta^{(1)}_{mn}|^2+(\nu_n-\nu_m)|\alpha^{(1)}_{mn}|^2\right]\nonumber\\
B_n=&\sum_{m\in C}\left(\nu_m \alpha^{(1)*}_{mn} \beta^{(1)}_{mn}\right)+\nu_n \alpha^{(0)*}_{nn} \beta^{(2)}_{nn}.
\end{align}	
}
Notice that \eqref{final:single:mode:reduced:state} can be written as 
\begin{align}\label{single:mode:decomposition}
\boldsymbol{\sigma}_{H,n}(h)=\boldsymbol{S}^{\dag}_{SMS}(h)\,\boldsymbol{\sigma}_{th,n}\,\boldsymbol{S}_{SMS}(h),
\end{align}
where the thermal state $\boldsymbol{\sigma}_{th,n}$ reads $\boldsymbol{\sigma}_{th,n}=(\nu_n+2\,A_n\,h^2)\,\mathds{1}_{2\times2}$ and the single mode squeezing matrix $\boldsymbol{S}_{SMS}(h)$ reads
\begin{align}
\boldsymbol{S}_{SMS}(h)=
	\begin{pmatrix}
	1 & \frac{B_n}{\nu_n}\,h^2\\
	\frac{B^*_n}{\nu_n}\,h^2 & 1
	\end{pmatrix}.
\end{align}
Single mode squeezing is an active transformation. Therefore, $\boldsymbol{S}^{\dag}_{SMS}(h)=\boldsymbol{S}_{SMS}(h)$ and we have 
\begin{align}
\boldsymbol{\sigma}_{H,n}(h)=(\nu_n+2\,A_n\,h^2)\,\boldsymbol{S}_{SMS}^2(h).
\end{align}
The decomposition \eqref{single:mode:decomposition} shows us that the final single-mode reduced state $\boldsymbol{\sigma}_{H,n}(h)$ is locally equivalent to the thermal state $\boldsymbol{\sigma}_{th,n}$ with a slightly modified temperature $T_n=T+\delta T_n\,h^2$, up to a local squeezing transformation. Here, $T$ is the initial temperature before the squeezing of the cavity and $\delta T_n h^2$ is the small change in local temperature with the expression
	\begin{align}\label{change:in:local:temperature}
	\frac{\delta T_n}{T}=&A_n\,\frac{2\,k_\text{B}\,T}{\hbar\,\omega_n}\,\sinh^2\left(\frac{\hbar\,\omega_n}{2\,k_\text{B}\,T}\right).
	\end{align}
Notice that the single mode squeezing operator $\boldsymbol{S}_{SMS}(h)$ has not changed the average population number of the mode, i.e., the state $\boldsymbol{\sigma}_{H,n}(h)$ has the same average number of excitations of the thermal state $(\nu_n+2\,A_n\,h^2)\,\mathds{1}_{2\times2}$. Furthermore, notice that this formula is correct in our perturbative regime if $\frac{\delta T_n}{T}\,h^2\ll1$. Finally, it cannot be applied in a straightforward fashion to $T=0$. This occurs because the inverse of the first derivative of the function $\coth x$ diverges exponentially for $x\rightarrow+\infty$. We also notice that for low temperatures we have $\delta T_n\sim A_n\,\frac{\hbar\,\omega_n}{2\,k_\text{B}}$. 

	\item[iii)] \textit{Work extraction}: The modes of cavity H are now initialised in thermal states. Cavities R and H and the battery B are coupled through an Hamiltonian with interaction term 
	\begin{align}
	H_I=r^{\dag}\,h^{\dag}\,b+r\,h\,b^{\dag},
	\end{align}
	where $r$ is the annihilation operator of  a chose mode $n$ in cavity R, $h$ is the annihilation operator of mode $n'$ in cavity H and $b$ is the annihilation operator of the battery harmonic oscillator B. We choose modes $n$ in R and $n'$ in H to be accessible by experimental means and to optimise energy influx. This interaction can be engineered in experimental setups and its role in quantum thermodynamical processes has been analysed in detail \cite{Goold:Huber:2016,Mitchison:Huber:2016}.
	
	The interaction Hamiltonian $H_I$ is not quadratic preventing us from exploiting Gaussian state formalism and covariance matrix techniques to obtain analytical results. However, we can provide an upper bound to the performance of this protocol, which can be achieved only in an idealised situation. 
	\item[iv)] \textit{Bounds on performance}: We now proceed and discuss the optimal bound of performance of this cycle. The cycle we described can be easily understood in terms of standard thermodynamics. In particular, given the temperature of the reservoir R and the final higher local temperature in the cavity H, we know that the efficiency $\eta_\text{cyc}$ of each cycle is always upper bounded by the Carnot efficiency $1-\frac{T}{T_n}=\frac{\delta T_n}{T}\,h^2$. We can use the change in local temperature (\ref{change:in:local:temperature}) to find an explicit expression for the highest possible efficiency as
	\begin{align}
	\eta_\text{cyc}=&A_n\,\frac{2\,k_\text{B}\,T}{\hbar\,\omega_n}\,\sinh^2\left(\frac{\hbar\,\omega_n}{2\,k_\text{B}\,T}\right)\,h^2.
	\end{align}
	The work (or free energy) extractible from mode $n$ per cycle is $W_c$. We have already computed the total amount of work $W$ that can ideally extracted per cycle in the system S. This is given by the sum of all extractible energy contributions $W_n$ per each mode. Notice that when the system S is composed of only one mode then $W_n\equiv W$. We have already introduced $W_n$ as $W_n=E_{H,n}-E_{R,n}$, where $E_{H,n}$ and $E_{R,n}$ are energies of the cavity H and reservoir R respectively.
	The work per cycle $W_c$ that can be extracted is therefore
	\begin{align}
	W_c\leq\eta_\text{cyc}\,W_n.
	\end{align}
	We notice that, as done throughout this work, we can express $W_n$ in series of $h$ and we find $W_n=W^{(2)}_n\,h^2+\mathcal{O}(h^4)$. Therefore, we have
	\begin{align}
	W^{(2)}_n=A_n.
	\end{align}
	 Putting all together one obtains the final bound on the amount of work $W_c$ that can be extracted per cycle. This reads
	\begin{align}\label{bound:on:energy:storage}
	\frac{W_c}{k_B\,T}\leq A_n^2\,\frac{2\,k_\text{B}\,T}{\hbar\,\omega_n}\,\sinh^2\left(\frac{\hbar\,\omega_n}{2\,k_\text{B}\,T}\right)\,h^4.
	\end{align}
\end{itemize}
It is clear that one of the major limiting factors for the practical implementation of this protocol is the total time each cycle takes. In particular, the total time is given by the sum of the individual times of each process: the time it takes to initialise the two cavities and the battery; the time it takes to excite the cavity H; the time it takes to couple the cavities to the battery in order extract the energy.

\subsection{Charging a quantum battery through motion and gravity: implementations and applications}\label{subsection:charging}
We notice that, although we can compute an explicit upper bound, the amount of extractible energy that can be obtained with this protocol seems much smaller than the available energy. One could be tempted to conclude that the above protocol is only good as an illustrative example of how to apply the results of this work, and that it is necessary to investigate better methods for energy extraction, if this energy is to be used for any practical purpose.

We now proceed to analyse the general behavior on the upper bound of the amount of storable energy $W_c$ with a cycle-like protocol. The amount of energy $W_n$ that can be extracted by the relativistic degree of freedom labeled by $n$ is always of the order $\mathcal{O}(h^2)$. This means that, regardless of the protocol, in each cycle we can transfer and store (much) less energy than the amount $W_n$ of available energy. In order to see that this statement is true, we look at the general efficiency in a thermodynamical cycle. The general efficiency $\eta$ is always bounded by the Carnot efficiency $\eta_\text{cyc}=1-\frac{T}{T_n}=\mathcal{O}(h^2)$. This, again, is a general property of the Carnot efficiency in this kind of protocol and it does not depend on the specific details of the protocol. The extractible energy $W_n$ is, as well, at most of second order, i.e., $W_n=\mathcal{O}(h^2)$, in any protocol that starts from an initial thermal state. The storable energy per cycle $W_c$ is therefore at most of fourth order, i.e., $W_c=\eta_\text{cyc}\,W_n=\mathcal{O}(h^4)$. 

However daunting the bound above might seem, we now proceed to show that one can still obtain non-negligible energy storage in a general protocol. We start by noting that we cannot analyse the previous cycle for $T=0$. However, we can look at extremely low temperatures, i.e., $\frac{\hbar\,\omega_n}{2\,k_\text{B}\,T}\ll1$. This implies that $\sinh(\frac{\hbar\,\omega_n}{2\,k_\text{B}\,T})\sim \exp[\frac{\hbar\,\omega_n}{2\,k_\text{B}\,T}]\gg1$. We can look at this regime as long as $\exp[\frac{\hbar\,\omega_n}{k_\text{B}\,T}]\,h^2\ll1$. We then note that this regime can still imply $\exp[\frac{\hbar\,\omega_n}{k_\text{B}\,T}]\,h^2\sim 5-10$\%, which in turn would imply that one could obtain a bound $\frac{W_c}{k_B\,T}\leq 1-10$\%, still within the validity of this perturbative regime.

Let us analyse when such a situation can occur {(a more detailed analysis of similar systems can be found in \cite{Sabin:Bruschi:2014})}. We can focus on phononic cavities in BEC-based technologies. Here we can tune the frequencies $\omega_n$ by tuning the length of the BEC or the interaction length. We can choose, for example, $\omega_1=200$Hz. The temperatures achievable in a BEC in the laboratory are typically of the order of $T=10$nK. Altogether, this implies $\frac{\hbar\,\omega_n}{2\,k_\text{B}\,T}\sim 1$. If we choose $\omega_1=1$kHz then we would get $\frac{\hbar\,\omega_n}{2\,k_\text{B}\,T}\sim 5$. These numbers lead to $\exp[\frac{\hbar\,\omega_n}{k_\text{B}\,T}]\sim 7.4$ and $\exp[\frac{\hbar\,\omega_n}{2\,k_\text{B}\,T}]\sim5\times 3.2 10^{6}$ respectively, which are much larger than unity. Notice that the range of frequencies that we are considering is easily obtainable in experimental setups based on BECs. This range and needs to be compatible with $h\sim10^{-3}-10^{-6}$ for BECs of the size of $L=10^{-6}$m, phononic speed of sound of $c_\textrm{S}=10^{-2}$m/s and accelerations of $0.1$g or lower. 
This combination and range of parameters can be obtained with current technology, which shows that BEC-based technologies are promising candidates for relativistic and quantum information tasks.

In contrast, optical cavities are typically characterised by frequencies of the order of $\omega_1=10^{14}$Hz and a perturbative parameter $h\sim10^{-19}$, when the cavity length is $L=1$cm and the acceleration is of the order of $10$g. This implies that one can tolerate extremely high temperatures before these approximations break down. However, this also implies that in order for the amount of stored energy to be significant, one needs either extremely large cavity accelerations or extremely large temperatures in the cavity. These regimes are currently out of experimental reach.

We conclude with the following remarks. First, in the scenarios considered above, it could be convenient to also look at multimode systems for energy extraction, where locally (i.e., in one-mode energy extraction cycles) one can witness violations of the Carnot bound. This might lead to considerably higher amount of energy to be stored in the battery. Second, we notice that the storable energy per cycle is small however, one can conceivably devise an extraction mechanism that can be repeated. This can occur, for example, if a monochromatic gravitational wave impinges on a trapped BEC. In that case, the battery could be charged with very small amounts of energy per cycle, but this procedure could be repeated enough times in order to obtain a measurable amount of energy. We leave it to future work to explore such opportunities.

\section{Results: Thermodynamic fluctuations in the cavity}
{In this section we are concerned with the possibility of thermodynamic fluctuations emerging in our cavity. This is not only of fundamental interest, but studying the behaviour of such thermodynamical fluctuations may help inform future concrete developments of thermal machines bases on confined relativistic quantum fields.} 

In order to study the role of {the} thermodynamic fluctuations in the cavity system we must derive the first and second moments of the work distribution following a perturbation of its Hamiltonian, see \ref{appen:peturbative mergence of fluctuations} for the full derivation. For the case of a system $S$ formed of one mode $p$. The moments are, respectively 
\begin{align}
G\left[\psi\right]=&\sum_{n_p} p_n E_{n_p}^{f(1)}h+\sum_{n_p} p_n E_{n_p}^{f(2)}h^2\nonumber\\ 
&+\sum_{m_q,n_p}p_n p_{m|n}^{(2)}\left(E_{m_q}^{f(0)}-E_{n_p}^i\right)h^2\label{equ:1moment}\\
G\left[\psi^2\right]=&\sum_{n_p}p_n\left(E_{n_p}^{f(1)}\right)^2h^2\nonumber\\
&+\sum_{m_q,n_p}p_np_{m|n}^{(2)}\left(E_{m_q}^{f(0)}-E_{n_p}^i\right)^2h^2.\label{equ:2moment}
\end{align}
Explicit expressions for these quantities are then computed in \ref{appen:peturbative mergence of fluctuations}.
For brevity we leave the full expressions \eqref{equ:full first moment} and \eqref{equ:full var of dist} out of the
main text, but continue to state the conclusions here.

In terms of the physical implications of these equations, it can be seen that only ``free falling'' boundaries under motion or dynamics of background spacetimes, such as those studied here  \cite{Barbado:Baez-Camargo:2018}  contribute to the variance of the work distribution. This variance disappears when one considers rigid boundary conditions such as those of an impinging gravitational wave \cite{Bruschi:Louko:2013}.  

In addition, by studying the form of equation \eqref{equ:1moment} and \eqref{equ:2moment} we see that the average amount of work extractable from the cavity $G[\psi]$, has a leading term which is first order in our perturbation parameter $h$, whereas the variance in the work distribution $G[\psi^2]$, only contains terms of order $h^2$. This shows that with some $h<<1$ the quantities of work that one could extract during a work extraction protocol are reasonably well clustered around the mean and can therefore be extracted with some certainty. 

\begin{figure}[h!]
	\centering
	\fbox{\includegraphics[scale=0.45]{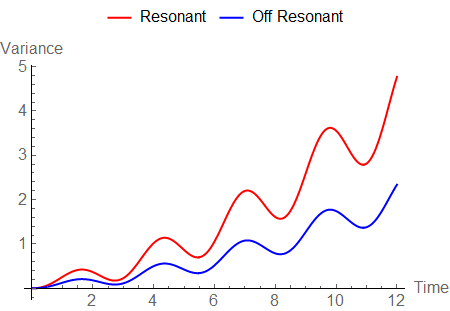}}
	\caption{Graph of the variance of the probability work distribution against time for an initially unpopulated cavity undergoing a perturbation of its surrounding spacetime due to an incoming gravitational wave \eqref{equ:var vacuum}. The Bogoliubov coefficients used can be found in equation \eqref{bogoliubov:coefficient}, with the corresponding resonant condition being $2\omega_k=\omega_{GW}$. {Parameters chosen for the cavity system and gravitational perturbation are of the order discussed in section \ref{subsection:charging}.}  \label{fig:protocol}}
\end{figure}

By specifying the initial state of the cavity and the gravitational effect responsible for the perturbation of its surrounding spacetime we can plot both the average and variance of the work distribution of the cavity. In Fig \ref{fig:protocol} we show the effect on the variance of the work distribution for a initially unpopulated cavity being perturbed by an incoming gravitational wave.   

We can see that the rate at which the variance of the work distribution grows is far greater when the cavity mode is in resonance with the gravitational wave. We can also see that the variance of the work distribution oscillates with a set frequency, which from an inspection of \eqref{equ:var vacuum} is dependant upon the frequency of the cavity mode. This demonstrates that with fine tuning of the cavity parameters, one can extract a set amount work from the cavity with a greater degree of certainty.

{In this work we have specialized our analysis to closed Gaussian dynamics, which is a natural choice for a model of an ideal thermal machine based on quantum fields confined in an ideal cavity. This approach includes most studies that focus on the phononic field of a BEC \cite{dalfovo1999theory}, or on the time evolution of an optomechanical system \cite{aspelmeyer2014cavity}. In such systems, the initial state of the field is naturally Gaussian (i.e., a thermal state), and it can be shown that the evolution is driven by a Hamiltonian that is quadratic to good approximation. For these reasons, we have restricted ourselves to the more manageable setup of Gaussian states evolving under ideal quadratic Hamiltonians. In order to extend the study of fluctuation relations beyond such closed Gaussian dynamics it is necessary instead to consider the evolution of the system coupled with an environment \cite{talkner2009fluctuation} to model decoherence, phonon loss, and other higher order effects, which is beyond the scope of our purpose here. We leave this to future work.}
\section{Applications: BEC-based gravitational wave detectors}\label{gravitational:wave:application}
We  are now able to proceed and show how our theoretical tools can be applied to a scenario of great interest. In this section we propose an application of our techniques, based on a scenario where the phononic modes of a BEC are affected by an impinging gravitational wave. These computations can be extended to cavities whose boundaries oscillate periodically. In particular, we are interested in, and will focus on, understanding how much energy can be extracted from the gravitational wave, and the amount that can be potentially stored.

\subsection{Impinging gravitational waves}
Previous work has investigated the details of cavity travel scenarios, where boundaries undergo arbitrary trajectories \cite{Bruschi:Louko:2013}, tailored specifically to different applications, such as quantum information processing \cite{Bruschi:Sabin:2016}. Here, we will focus on the applications for detection of gravitational waves.

A gravitational wave is small a perturbation of the metric predicted by the theory of general relativity and typically emitted by large accelerating masses, pulsating asymmetric masses or very energetic astrophysical events \cite{Abbott:Abbott:2016}. When impinging a distant object, like an antenna on the Earth, the wave ``stretches and compresses'' spacetime, therefore affecting the proper distance between any two points \cite{Abbott:Abbott:2016}. This fact can be used, for example, in laser-interferometers to measure the change in length of one of the two arms of the detector, induced by an impinging gravitational wave, which are traveled by laser pulses \cite{Abbott:Abbott:2016}. There are different such ``antennas'' that are operative and are planned for the future, such as aLIGO and aVIRGO \cite{Abbott:2016}. The main difficulties lie in the low strength of signal, compared to the background noise, in the expected frequency domains and the randomness of the events. 

Recently, a scheme was proposed to detect gravitational waves with micrometer quantum systems, known as BECs \cite{Sabin:Bruschi:2014}. It was shown that a trapped BEC with small phononic excitations can be effectively modelled as a massless bosonic field $\hat \Phi$ which moves slower than light (at the speed of sound $c_\textrm{S}\ll c$) and is trapped in a cavity \cite{Dalfovo:Giorgini:1999}. A gravitational wave that passes through the system affects the phononic modes of the cavity \cite{Sabin:Bruschi:2014}. This phenomenon induces detectable changes in the quantum state of the phonons that can be used as a signature of the metric perturbation with current technology \cite{Sabin:Bruschi:2014}. This process requires energy to be transferred from the gravitational perturbation to the quantum field. Here we will analyse the performance of this process.

\subsection{Modelling gravitational waves}
The model of a field contained in a cavity that we have presented in this work applies in a straightforward fashion to this new setup, except that the speed of light $c$ is replaced with the speed of sound $c_\textrm{S}$, see \cite{Sabin:Bruschi:2014}. We also assume that the BEC is \textit{$3$-dimensional}, it is strongly confined in two dimensions and elongated in the $x$ direction. In addition, we will consider two scenarios: i) the case where the boundaries are rigid. ii) The  case where the boundaries ``free fall'', as in the case of LIGO \cite{Abbott:Abbott:2016}. In the former we can model the effect of a sinusoidal impinging gravitational wave as a time dependent cavity length $L_x(t)=L_x(1+h\,\sin(\omega_\text{GW}\,t))$ for time $t$, where $h\ll1$ is the amplitude of the wave and $\omega_\text{GW}$ is its frequency \cite{Bruschi:Louko:2013,Abbott:Abbott:2016}. 

Then, the relevant Bogoliubov coefficients that are in resonance with the gravitational wave read
\begin{align}\label{bogoliubov:coefficient}
\beta_{klm,\tilde{k}lm}=&\pi^2\,c_\text{s}^4\,\frac{k\,\tilde{k}}{L_x^4\,\omega_\text{GW}^2\sqrt{\omega_{klm}\,\omega_{\tilde{k}lm}}}\,t\nonumber\\
\beta_{klm,klm}=&\pi^2\,c_\text{s}^2\,\frac{l^2/L_y^2-m^2/L_z^2}{2\,\omega_\text{GW}}\,t,
\end{align}
where $k,l,m$ are the quantum numbers associated to the three standard Cartesian axes, the trap has dimensions $L_x,L_y,L_z$, and the frequency $\omega_{klm}$ reads $\omega_{klm}^2:=\pi^2\,c_\text{s}^2\,(k^2/L_x^2+l^2/L_y^2+m^2/L_z^2)$. Here $c_\text{s}$ is the speed of sound of the phonons in the BEC. The resonance conditions $\omega_{klm}+\omega_{\tilde{k}ml}=\omega_\text{GW}$ and $2\,\omega_{klm}=\omega_\text{GW}$ respectively give a significant effect after a sufficiently long time $t$, and in this case the coefficients \eqref{bogoliubov:coefficient} are the only one that contribute effectively to the population of the cavity modes.

Notice that below we use the label $k\equiv(k,l,m)$, $\tilde{k}\equiv(\tilde{k},l,m)$ and $k_\text{E}\equiv(k_\text{E},l_\text{E},m_\text{E})$, and that the coefficients \eqref{bogoliubov:coefficient} reduce to the ones found in the literature for $1+1$ dimensional systems \cite{Sabin:Bruschi:2014}.

\subsection{Energy transfer from gravitational waves: initial single mode squeezed state}
We would like to study the efficiency of energy transfer from the gravitational wave given an initial single mode squeezed states for the two scenarios discussed so far. We assume that {the} system S is composed of modes $k$ and $\tilde{k}$, which are in resonance with the gravitational wave (i.e., $\omega_{klm}+\omega_{\tilde{k}ml}=\omega_\text{GW}$) in the case of rigid boundaries, or one mode $k$ resonant to the gravitational wave through $2\,\omega_{klm}=\omega_\text{GW}$ for the case of free-falling boundaries. The modes are initially in a two-mode squeezed thermal state with squeezing $r$, and a single mode squeezed state with squeezing $s$ respectively. After a sufficiently long time $t$, the efficiency has the expression 
\begin{align}\label{ascecec}
\frac{1}{\eta}=&1+\frac{\sum_{k_\text{E}\in E} k_\text{E}\left[\left|\Lambda_{k_\text{E}k}(t,s)\right|^2+\left|\Lambda_{k_\text{E}\tilde{k}}(t,s)\right|^2\right]}{\pi^2\,c_\text{s}^4\,\frac{k\,\tilde{k}}{L_x^4\,\omega_\text{GW}^2\sqrt{\omega_{klm}\,\omega_{\tilde{k}lm}}}\,\tanh s}\frac{1}{t}h\nonumber\\
\frac{1}{\eta}=&1+\frac{\sum_{k_\text{E}\in E}\,k_\text{E}\,\left|\Lambda_{k_\text{E}k}(t,s)\right|^2}{\pi^2\,c_\text{s}^2\,\frac{l^2/L_y^2-m^2/L_z^2}{2\,\omega_\text{GW}}\,\tanh s}\frac{1}{t}h,
\end{align}
where we have introduced
\begin{align*}
\Lambda_{k_\text{E}k}(t,s):=\beta_{k_\text{E}k}^{(1)*}+\alpha_{k_\text{E}k}^{(1)}\,\tanh s+\frac{1}{2}\sum_{p\in E}\mathcal{Z}_{k_\text{E}p}(t)
\end{align*} 
for convenience of presentation.

We notice from \eqref{ascecec} that in both cases the efficiency approaches unity with an error of the order $\mathcal{O}(\frac{1}{t})$. Initial squeezing $s$ helps in both cases, that is, it increases the fidelity with exponential gain.

\subsection{Energy transfer and leakage}
We have computed the efficiency \eqref{ascecec} for the case of a sinusoidal gravitational wave impinging on a (effectively one-dimensional) BEC gravitational wave detector. It is evident from \eqref{ascecec} that the efficiency increases with the increase of time. This can be easily explained by the fact that the majority of the energy delivered by the gravitational wave is transferred to, and absorbed by, the resonant modes. The efficiency \eqref{ascecec} for this scenario is the result of a series of approximations and does not apply for all times, but only as long as $t\,h\ll1$.

We note that that the behaviour of the efficiency \eqref{ascecec} would dramatically change if, for example, the two resonant modes were one in the system S and one in the environment E, i.e., a scenario where the control over the system S is not very good and part of the energy ``leaks'' to the environment E.

{\subsection{Applicability} 
The aim of our work is to provide theoretical connections between quantum thermodynamics and relativistic quantum fields, which are of fundamental interest. Nevertheless, we would like to add a few remarks on the experimental feasibility of the scheme proposed to detect gravitational perturbations with a trapped BEC. Among all existing spacetime perturbations, of particular interest are long lived monocromatic gravitational waves produced by neutron stars \cite{Andersson2004Gravitational,Ju2000Detection}, as well as spacetime perturbations produced by light bosonic dark matter candidates such as light pseudo-Nambu–Goldstones (axions and “axion-like particles”—ALPs) \cite{Frigerio2011Sub}, massive hidden vector bosons (dark photons) \cite{Redondo2009Massive}, and light scalars (Moduli/dilatons) \cite{Cicoli2011Anisotropic,Damour1994string}. Most neutron star models predict that the strain produced gravitational waves is very weak (below $10^{-20}$, see \cite{Ju2000Detection}), due to the enormous distances between the Earth and their sources. However, in the past decade there have been significant advances in precision measurements of physical parameters in quantum experiments fuelled by greater control on quantum systems. These advances inspired the proposal to use BECs to detect gravitational waves \cite{Sabin2014Phonon}, as well as other small systems such as levitated optomechanical devices \cite{Arvanitaki2013Detecting}.

The sensitivity of the phononic gravitational wave detector \cite{Sabin2014Phonon} relies upon a number of mutually contingent parameters: the size of the BEC, its shape, the frequencies of the two phonon modes that are initially squeezed, the number of atoms in the BEC, the number of phonons being squeezed, the measurement time, the atom-atom interactions, and the frequency of the signal being detected.  These parameters can be optimised simultaneously under non-linear constraints, most of which stem from physical considerations relating the parameters to one another and considering different sources of noise, such as thermal noise \cite{Sabin2016Thermal}, decoherence due to phonon-phonon interactions \cite{howl2017quantum} and three body recombination. All of the parameters are subject to distinct bounds, which can be drawn from state-of-the-art experiments. The parameters and sensitivity stay completely within the bounds of existing technological capabilities \cite{Howl2020Quantum}. As an example of the quantities involved for a single sodium condensate with $10^8$ atoms and integration time $\tau = 1$ year, gravitational waves between a few kHz and $100$kHz can be in principle detected with sensitivities ranging between $5\times10^{-20}$ and $5\times10^{-23}$. The high frequency range can be of interest in the search of light dark matter candidates. 

A more accessible proof-of-principle experiment has been proposed to study the dynamical response of a BEC to a small gravitational field engineered in the laboratory using an oscillating bead \cite{ratzel2018dynamical}. The aim of that study was to propose an experiment to demonstrate the key prediction of quantum field theory in curved spacetime: that spacetime dynamics can produce excitations in quantum fields. The proposal considers various sources of noise and the estimates show that gravitational perturbations generated by a $1$g titanium bead oscillating at a few Hertz and placed at millimeter distances from a sodium BEC (containing approximately $10^8$ atoms) produce one phonon every 10 seconds when the phononic field is initially squeezed. Although phonons are typically short lived, it has been also shown that it is possible to maintain phonons in a trapped BEC for a few seconds, a period of time that is sufficiently long for the system to interact with the background curvature perturbations \cite{howl2017quantum}.

Our study is a first attempt to understand quantum fields in flat or curved spacetime through the lens of quantum thermodynamics. In addition, it also informs us on the fundamental aspects of physical systems that exhibit relativistic and quantum features. In order to highlight the importance of this line of research, we have applied our results to a model of a confined quantum field (i.e., the phononic field of a trapped BEC) that is subject to spacetime distortions and can be used to probe such perturbations. BEC-based experimental setups can now be found in most cold-atom laboratories, where they can be manipulated and measured with extremely great precision. This makes them top candidates for experimental studies of fundamental science at the overlap of relativity and quantum mechanics. The promising studies mentioned above have motivated novel theoretical research and experimental proposals based on light interacting with matter, or quantum fields trapped into cavities. Our work falls within this last approach.} 
\section{Conclusions}\label{conclusions}

In this work we have introduced techniques to understand the performance of localised relativistic quantum fields, such as the electromagnetic field confined in moving cavities or the phononic modes of a trapped BEC, as ``machines'' that can ``extract'' energy from relativistic degrees of freedom. These include motion of boundaries or dynamical changes in the background spacetime. Each localised setup is divided into a system S, which can be accessed, manipulated and controlled, and an environment E, which acts as a reservoir of energy and entropy. We have obtained analytical expressions for the efficiency of transfer of energy into the system (i.e., field excitations) from the classical change of the boundaries of the cavity which can be induced by motion, or from a classical gravitational wave. The latter scenario is of great relevance for modern applications within gravitational wave science, astrophysics and cosmology \cite{Sathyaprakash:Schutz:2009}. In this context we have shown that sinusoidal modulations, which can be induced by impinging gravitational waves, are associated to an efficiency that grows with the duration of the signal itself and approaches unity in realistic ultra-cold setups. Furthermore, we have provided an analysis of the distribution of the work absorbed by the system using modern results derived from fluctuation theorems. All together our outcomes indicate that energy extraction is efficient in cavity setups of the type studied here. 

These exciting results suggest that it can be of fundamental and technological relevance to develop technologies tailored at extraction of energy from classical degrees of freedom, relativistic degrees of freedom and gravitational fields. In particular, our work aids the analysis of feasibility of recently proposed technologies for the detection of gravitational wave by BEC-based antennas \cite{Sabin:Bruschi:2014}. Furthermore, we believe that this work provides the foundation for a novel way to characterize localized quantum systems as (relativistic and quantum) thermal machines. Extension of this work to include loss and decoherence, and optimzed shapes tailored for detection of signals from realistic sources, such as binary neutron star mergers and pulsars in the case of gravitational drive, is left to future work.

\section*{Acknowledgments}
We thank the late Jacob Bekenstein, Tsvi Piran, Jandu Dradouma, Gabriel Landi and Luis Cort\'es Barbado for useful comments and suggestions. We are greatly indebted to Marcus Huber for his help with the thermodynamics aspects of this work.
D.E.B. was initially supported by the I-CORE Program of the Planning and Budgeting Committee and the Israel Science Foundation (grant No. 1937/12), as well as by the Israel Science Foundation personal grant No. 24/12. D.E.B. was also supported by John Templeton Foundation grant no. 58745 and the COST Actions MP1209 and MP1304. D.E.B. acknowledges support from the LIQUID collaboration, in which context part of this work was done. D. E. B. also acknowledges later hospitality from the Hebrew University of Jerusalem and the University of Vienna, where part of this work was done. 
B.M. acknowledges the financial support of the EPSRC (Grant No. EP/N50970X/1).


\bibliographystyle{elsarticle-num}\biboptions{sort&compress}
\bibliography{ThermalCavitiesBiblio}


\onecolumn
\newpage
\appendix

\section{Properties of perturbative symplectic (Bogoliubov) transformations and covariance matrices}\label{appendix:zero}
Here we provide all of the necessary information concerning the properties and expression of symplectic (or Bogoliubov) transformations in the perturbative regime, including expression and properties of covariance matrices subject to such transformations and the symplectic eigenvalues. The list below is exhaustive. 

\subsection{Bogoliubov (symplectic) transformations}\label{appendix:one:Bogo}
A symplectic matrix $\boldsymbol{S}(\lambda)$ satisfies the symplectic identity $\boldsymbol{S}\,\boldsymbol{\Omega}\,\boldsymbol{S}^\dag=\boldsymbol{S}^\dag\,\boldsymbol{\Omega}\,\boldsymbol{S}=\boldsymbol{\Omega}$, where $\boldsymbol{\Omega}$ is the symplectic form. In our basis, the symplectic form is diagonal and has the expression $\boldsymbol{\Omega}=\text{diag}(-i,-i,\ldots,i,i,\ldots)$, and satisfies $\boldsymbol{\Omega}^2=-\mathds{1}$ and $\boldsymbol{\Omega}\,\boldsymbol{\Omega}^\dag=\mathds{1}$. A symplectic matrix $\boldsymbol{S}(\lambda)$ has the general expression 
\begin{align}\label{generic:symplectic:matrix}
\boldsymbol{S}(\lambda)=\overset{\leftarrow}{\mathcal{T}}\exp\left[\frac{1}{\hbar}\boldsymbol{\Omega}\,\int_0^\lambda\,d\lambda'\,\boldsymbol{H}(\lambda')\right],
\end{align}
where $\boldsymbol{H}(\lambda)$ is an appropriate Hermitian matrix that might depend explicitly on $\lambda$ and not commute with itself for different values of $\lambda$. An example of such matrix is a (time-dependent) Hamiltonian, with $\lambda\equiv t$.\\
We note here that a symplectic matrix $\boldsymbol{S}(\lambda)$ represents a unitary operator obtained from a (time-ordered) exponential of an operator $\hat{H}(\lambda)$ that is quadratic in the creation and annihilation operators, and which defines $\boldsymbol{H}$ through $\hat{H}=\frac{1}{2}\,\mathbb{X}^\dag\,\boldsymbol{H}\,\mathbb{X}$.

The symplectic matrix $\boldsymbol{S}(\lambda)$ has the general expression
\begin{align}
\boldsymbol{S}=
\begin{pmatrix}
\boldsymbol{\alpha} & \boldsymbol{\beta}\\
\boldsymbol{\beta}^* & \boldsymbol{\alpha}^*
\end{pmatrix},
\end{align}
where the square matrices $\boldsymbol{\alpha}$ and $\boldsymbol{\beta}$ collect the well-known Bogoliubov coefficients $\alpha_{nm}$ and $\beta_{nm}$.

The Bogoliubov identities $\boldsymbol{\alpha}\,\boldsymbol{\alpha}^{\dag}-\boldsymbol{\beta}\,\boldsymbol{\beta}^{\dag}=\mathds{1}$ and $\boldsymbol{\alpha}\,\boldsymbol{\beta}^{\text{T}}-\boldsymbol{\beta}\,\boldsymbol{\alpha}^{\text{T}}=0$ can be obtained as a direct consequence of the symplectic identity $\boldsymbol{S}\,\boldsymbol{\Omega}\,\boldsymbol{S}^\dag=\boldsymbol{\Omega}$. Notice that the identity $\boldsymbol{S}^\dag\,\boldsymbol{\Omega}\,\boldsymbol{S}=\boldsymbol{\Omega}$ gives an equivalent version of the Bogoliubov identities, namely $\boldsymbol{\alpha}^{\dag}\,\boldsymbol{\alpha}-\boldsymbol{\beta}^{\text{T}}\,\boldsymbol{\beta}^{*}=\mathds{1}$ and $\boldsymbol{\alpha}^{\dag}\,\boldsymbol{\beta}-\boldsymbol{\beta}^{\text{T}}\,\boldsymbol{\alpha}^{*}=0$. 

As a consequence of the symplectic identity, we also have that
\begin{align}
\boldsymbol{S}^{-1}=&-\boldsymbol{\Omega}\,\boldsymbol{S}^\dag\,\boldsymbol{\Omega}\nonumber\\
\boldsymbol{S}^{-1\dag}=&-\boldsymbol{\Omega}\,\boldsymbol{S}\,\boldsymbol{\Omega}.
\end{align}

\subsection{Perturbative regime: Bogoliubov transformations}\label{appendix:one:perturbative:regime:Bogo}
We now decide to look at perturbative regimes of the all quantities when the symplectic matrix $\boldsymbol{S}(\lambda)$ depends on a small parameter $h\ll1$.

We start by the symplectic matrix $\boldsymbol{S}$, which has the perturbative expression
\begin{align}\label{perturbative:symplectic:identity:appendix}
\boldsymbol{S}^{(0)}\,\boldsymbol{\Omega}\,\boldsymbol{S}^{(0)\dag}=&\boldsymbol{\Omega}\nonumber\\
\boldsymbol{S}^{(0)}\,\boldsymbol{\Omega}\,\boldsymbol{S}^{(1)\dag}+\boldsymbol{S}^{(1)}\,\boldsymbol{\Omega}\,\boldsymbol{S}^{(0)\dag}=&0\nonumber\\
\boldsymbol{S}^{(0)}\,\boldsymbol{\Omega}\,\boldsymbol{S}^{(2)\dag}+\boldsymbol{S}^{(1)}\,\boldsymbol{\Omega}\,\boldsymbol{S}^{(1)\dag}+\boldsymbol{S}^{(2)}\,\boldsymbol{\Omega}\,\boldsymbol{S}^{(0)\dag}=&0.
\end{align}
Therefore, the perturbative Bogoliubov identities read
\begin{align}\label{perturbative:contributions:to:bogoliubov:identities}
\boldsymbol{\alpha}^{(0)}\,\boldsymbol{\alpha}^{(0)*}=\mathds{1}\nonumber\\
\boldsymbol{\alpha}^{(0)}\,\boldsymbol{\alpha}^{(1)\dag}+\boldsymbol{\alpha}^{(1)}\,\boldsymbol{\alpha}^{(0)*}=0\nonumber\\
\boldsymbol{\alpha}^{(0)}\,\boldsymbol{\alpha}^{(2)\dag}+\boldsymbol{\alpha}^{(2)}\,\boldsymbol{\alpha}^{(0)*}+\boldsymbol{\alpha}^{(1)}\,\boldsymbol{\alpha}^{(1)\dag}-\boldsymbol{\beta}^{(1)}\,\boldsymbol{\beta}^{(1)\dag}=0\nonumber\\
\boldsymbol{\alpha}^{(0)}\,\boldsymbol{\beta}^{(1)\text{T}}-\boldsymbol{\beta}^{(1)}\,\boldsymbol{\alpha}^{(0)}=0\nonumber\\
\boldsymbol{\alpha}^{(1)}\,\boldsymbol{\beta}^{(1)\text{T}}+\boldsymbol{\alpha}^{(0)}\,\boldsymbol{\beta}^{(2)\text{T}}-\boldsymbol{\beta}^{(1)}\,\boldsymbol{\alpha}^{(1)\text{T}}-\boldsymbol{\beta}^{(2)}\,\boldsymbol{\alpha}^{(0)}=0.
\end{align}
The equivalent Bogoliubov identities $\boldsymbol{\alpha}^{\dag}\,\boldsymbol{\alpha}-\boldsymbol{\beta}^{\text{T}}\,\boldsymbol{\beta}^{*}=\mathds{1}$ and $\boldsymbol{\alpha}^{\dag}\,\boldsymbol{\beta}-\boldsymbol{\beta}^{\text{T}}\,\boldsymbol{\alpha}^{*}=0$ have their own perturbative expansions analogous to \eqref{perturbative:contributions:to:bogoliubov:identities}, which we can be easily obtained in the same fashion and we omit here.

\subsection{Perturbative regime: Bogoliubov transformations induced by Hamiltonians}\label{appendix:one:perturbative:bogoliubov:hamiltonians}
We have noted that the generic symplectic matrix $\boldsymbol{S}(\lambda)$ has the expression \eqref{generic:symplectic:matrix}. If we consider the case where $\boldsymbol{H}(\lambda)$ represents a usual Hamiltonian, then we can write $\boldsymbol{H}(\lambda)=\boldsymbol{H}_0(\lambda)+\boldsymbol{H}_\text{I}(\lambda)$, where $\boldsymbol{H}_0$ is the matrix representation of the free Hamiltonian.

If one assumes that the perturbation is due to $\boldsymbol{H}_\text{I}(\lambda)$, and that $\boldsymbol{H}_0$ is independent of $\lambda$ to all orders of interest, then this is enough to imply that $\boldsymbol{S}^{(0)}(\lambda)=\exp[\frac{1}{\hbar}\boldsymbol{\Omega}\,\boldsymbol{H}_0\,t]=\exp[-i\,\boldsymbol{\omega}\,t]$, where $\boldsymbol{\omega}:=\text{diag}(\omega_1,\ldots,\omega_n,\ldots,\omega_1,\ldots,\omega_n,\ldots)$.

Note that we can therefore write
\begin{align}
\boldsymbol{S}(\lambda)=\boldsymbol{S}^{(0)}(\lambda)\,\overset{\leftarrow}{\mathcal{T}}\exp\left[\frac{1}{\hbar}\boldsymbol{\Omega}\,\int_0^\lambda\,d\lambda'\,\boldsymbol{S}^{(0)\dag}(\lambda')\,\boldsymbol{H}_\text{I}(\lambda')\,\boldsymbol{S}^{(0)}(\lambda')\right].
\end{align}
This expression will become useful later on.

\subsection{Perturbative regime: quantum state}\label{appendix:one:perturbative:regime:state}
We continue by discussing the perturbative expression of the quantum state. In this work we focus on Gaussian states \cite{Adesso:Ragy:2014}.

A Gaussian state is uniquely defined by its covariance matrix $\boldsymbol{\sigma}$ and the vector $\boldsymbol{d}$ of its first moments.
A covariance matrix $\boldsymbol{\sigma}$ can always be written as $\boldsymbol{\sigma}=\boldsymbol{S}\,\boldsymbol{\nu}_{\oplus}\,\boldsymbol{S}^\dag$, where the diagonal matrix $\boldsymbol{\nu}_{\oplus}:=\text{diag}(\nu_1,\nu_2,\ldots,\nu_1,\nu_2,\ldots)$ is called the \textit{Williamson form} of $\boldsymbol{\sigma}$ and the $\{\nu_n\}$ are the symplectic eigenvalues of $\boldsymbol{\sigma}$, i.e., the eigenvalues of $|i\,\boldsymbol{\Omega}\,\boldsymbol{\sigma}|$.
The determinant of a covariance matrix $\boldsymbol{\sigma}$ therefore reads $\text{det}(\boldsymbol{\sigma})=\prod_n\,\nu^2_n$.

The \textit{general} expression of a covariance matrix $\boldsymbol{\sigma}$ is
\begin{align}
\boldsymbol{\sigma}=
\begin{pmatrix}
\boldsymbol{U} & \boldsymbol{V}\\
\boldsymbol{V}^* & \boldsymbol{U}^*
\end{pmatrix},\label{arbitrary:state:decomposition}
\end{align}
where $\boldsymbol{U}=\boldsymbol{U}^{\dag}$ and $\boldsymbol{V}=\boldsymbol{V}^{\text{T}}$.

The final state $\boldsymbol{\sigma}_{\text{f}}=\boldsymbol{S}(\lambda)\,\boldsymbol{\sigma}_{\text{i}}\,\boldsymbol{S}^{\dag}(\lambda)$, after a linear transformation $\boldsymbol{S}(\lambda)$, reads
\begin{align}\label{big:final:state}
\boldsymbol{\sigma}_{\text{f}}=\begin{pmatrix}
 \boldsymbol{U}(\lambda) & \boldsymbol{V}(\lambda)\\
\boldsymbol{V}^*(\lambda) 
& \boldsymbol{U}^*(\lambda)
\end{pmatrix},
\end{align}
where, the matrices $\boldsymbol{U}(\lambda)$ and $\boldsymbol{V}(\lambda)$ have the expression
\begin{align}
\boldsymbol{U}(\lambda)=\boldsymbol{\alpha}\,\boldsymbol{U}_{\text{i}}\,\boldsymbol{\alpha}^{\dag}+\boldsymbol{\alpha}\,\boldsymbol{V}_{\text{i}}\,\boldsymbol{\beta}^{\dag}+\boldsymbol{\beta}\,\boldsymbol{V}_{\text{i}}^*\,\boldsymbol{\alpha}^{\dag}+\boldsymbol{\beta}\,\boldsymbol{U}_{\text{i}}^*\,\boldsymbol{\beta}^{\dag}\nonumber\\
\boldsymbol{V}(\lambda)=\boldsymbol{\alpha}\,\boldsymbol{U}_{\text{i}}\,\boldsymbol{\beta}^{\text{T}}+\boldsymbol{\alpha}\,\boldsymbol{V}_{\text{i}}\,\boldsymbol{\alpha}^{\text{T}}+\boldsymbol{\beta}\,\boldsymbol{V}_{\text{i}}^*\,\boldsymbol{\beta}^{\text{T}}+\boldsymbol{\beta}\,\boldsymbol{U}_{\text{i}}^*\,\boldsymbol{\alpha}^{\text{T}}
\end{align}
and the matrices $\boldsymbol{U}_{\text{i}},\boldsymbol{V}_{\text{i}}$ correspond to the corresponding expression \eqref{arbitrary:state:decomposition} of the initial state $\boldsymbol{\sigma}_{\text{i}}$.

Note that the Bogoliubov transformation $\boldsymbol{S}(\lambda)$ can represent any linear transformation including time evolution, in which case $\lambda\equiv t$.

We continue by expanding to lowest order the state of the system after a Bogoliubov transformation has applied parametrised by the small parameter $h$. We have
\begin{align}\label{big:final:state:perturbative:expansion}
\boldsymbol{\sigma}(\lambda)=&
\begin{pmatrix}
\boldsymbol{U}^{(0)}(\lambda)
& \boldsymbol{V}^{(0)}(\lambda)\\  
\boldsymbol{V}^{(0)*}(\lambda) 
& \boldsymbol{U}^{(0)*}(\lambda)
\end{pmatrix}
+
\begin{pmatrix}
\boldsymbol{U}^{(1)}(\lambda)
& \boldsymbol{V}^{(1)}(\lambda)\\  
\boldsymbol{V}^{(1)*}(\lambda) 
& \boldsymbol{U}^{(1)*}(\lambda)
\end{pmatrix}h
+
\begin{pmatrix}
\boldsymbol{U}^{(2)}(\lambda)
& \boldsymbol{V}^{(2)}(\lambda)\\  
\boldsymbol{V}^{(2)*}(\lambda) 
& \boldsymbol{U}^{(2)*}(\lambda)
\end{pmatrix}h^2+\mathcal{O}(h^3),
\end{align}
where we have introduced the zero order contributions
\begin{align}
\boldsymbol{U}^{(0)}(\lambda)=&\boldsymbol{\alpha}^{(0)}\,\boldsymbol{U}_{\text{i}}\,\boldsymbol{\alpha}^{(0)*} \nonumber\\
\boldsymbol{V}^{(0)}(\lambda)=&\boldsymbol{\alpha}^{(0)}\,\boldsymbol{V}_{\text{i}}\,\boldsymbol{\alpha}^{(0)}
\end{align}
and the first order contributions 
\begin{align}
\boldsymbol{U}^{(1)}(\lambda)=&\boldsymbol{\alpha}^{(1)}\,\boldsymbol{U}_{\text{i}}\,\boldsymbol{\alpha}^{(0)*}
+\boldsymbol{\alpha}^{(0)}\,\boldsymbol{U}_{\text{i}}\,\boldsymbol{\alpha}^{(1)\dag}+\boldsymbol{\alpha}^{(0)}\,\boldsymbol{V}_{\text{i}}\,\boldsymbol{\beta}^{(1)\dag}
+\boldsymbol{\beta}^{(1)}\,\boldsymbol{V}_{\text{i}}^*\,\boldsymbol{\alpha}^{(0)*}    \nonumber\\
\boldsymbol{V}^{(1)}(\lambda)=&\boldsymbol{\alpha}^{(1)}\,\boldsymbol{V}_{\text{i}}\,\boldsymbol{\alpha}^{(0)}
+\boldsymbol{\alpha}^{(0)}\,\boldsymbol{V}_{\text{i}}\,\boldsymbol{\alpha}^{(1)\text{T}}+\boldsymbol{\alpha}^{(0)}\,\boldsymbol{U}_{\text{i}}\,\boldsymbol{\beta}^{(1)\text{T}}
+\boldsymbol{\beta}^{(1)}\,\boldsymbol{U}_{\text{i}}^*\,\boldsymbol{\alpha}^{(0)}
\end{align}
for simplicity of presentation of the formula above.

We anticipate that the second order contribution $\boldsymbol{U}^{(2)}(\lambda)$ to the matrix $\boldsymbol{U}(\lambda)$ is the only relevant second order contribution to this work, and it reads
\begin{align}\label{k2}
\boldsymbol{U}^{(2)}(\lambda)=&\boldsymbol{\alpha}^{(0)}\,\boldsymbol{U}_{\text{i}}\,\boldsymbol{\alpha}^{(2)\dag}
+\boldsymbol{\alpha}^{(1)}\,\boldsymbol{U}_{\text{i}}\,\boldsymbol{\alpha}^{(1)\dag}+\boldsymbol{\alpha}^{(2)}\,\boldsymbol{U}_{\text{i}}\,\boldsymbol{\alpha}^{(0)*}+\boldsymbol{\beta}^{(1)}\,\boldsymbol{U}_{\text{i}}^*\,\boldsymbol{\beta}^{(1)\dag}+\boldsymbol{\alpha}^{(1)}\,\boldsymbol{V}_{\text{i}}\,\boldsymbol{\beta}^{(1)\dag}
+\boldsymbol{\alpha}^{(0)}\,\boldsymbol{V}_{\text{i}}\,\boldsymbol{\beta}^{(2)\dag}\nonumber\\
&+\boldsymbol{\beta}^{(1)}\,\boldsymbol{V}^*_{\text{i}}\,\boldsymbol{\alpha}^{(1)\dag}
+\boldsymbol{\beta}^{(2)}\,\boldsymbol{V}_{\text{i}}^*\,\boldsymbol{\alpha}^{(0)*}.
\end{align}

We end this important section with the following observation. Let $\boldsymbol{\sigma}(\lambda)$ be the covariance matrix representing a multimode bosonic quantum state. For simplicity, we can assume that there are a system S and an environment E both comprising many bosonic modes, as considered in this work. The reduced state $\boldsymbol{\sigma}_\textrm{S}(\lambda)$ of the system S can be easily computed from the full state $\boldsymbol{\sigma}(\lambda)$ by simply deleting the rows and columns corresponding to the modes of the environment E, see \cite{Adesso:Ragy:2014}.

In this case, the reduced state $\boldsymbol{\sigma}(\lambda)$ can be expressed as
\begin{align}\label{system:reduced:state:appendix}
\boldsymbol{\sigma}_\textrm{S}(\lambda)=\boldsymbol{s}(\lambda)\,\boldsymbol{\nu}_{\oplus,\textrm{S}}(\lambda)\,\boldsymbol{s}^\dag(\lambda),
\end{align}
where $\boldsymbol{s}(\lambda)$ is a symplectic matrix that is \textit{different} from the matrix $\boldsymbol{S}(\lambda)$ used to compute the final full state  $\boldsymbol{\sigma}(\lambda)=\boldsymbol{S}(\lambda)\,\boldsymbol{\sigma}_{\text{i}}\,\boldsymbol{S}^\dag(\lambda)$, and $\boldsymbol{\nu}_{\oplus,\textrm{S}}(\lambda)$ is the Williamson form of this reduced state of the system S. Note that $\boldsymbol{\nu}_{\oplus,\textrm{S}}(\lambda)$ depends potentially on $\lambda$.

\subsection{Perturbative regime: symplectic eigenvalues}\label{appendix:one:perturbative:regime:symplectic:eigenvalues}
We here study the perturbative expressions of the symplectic eigenvalues $\nu_n$. The matrix $i\,\boldsymbol{\Omega}\,\boldsymbol{\sigma}$ has spectrum $\{\pm\nu_n\}$. In particular, diagonalization gives us the matrix $\text{diag}(\nu_1,\nu_2,\ldots,-\nu_1,-\nu_2,\ldots)$.

The symplectic eigenvalues $\nu_n$ have the general expression $\nu_n=\mathrm{coth}(\frac{\hbar\,\omega_n}{k_{\text{b}}\,T_n})$ for an appropriate temperature $T_n$ and therefore satisfy $\nu_n\geq1$.
These eigenvalues have the general perturbative expression 
\begin{align}\label{perturbative:symplectic:eigenvalue:expression}
\nu_n=\nu_n^{(0)}+\nu_n^{(1)}\,h+\nu_n^{(2)}\,h^2+\mathcal{O}(h^4).
\end{align}
The contributions $\nu_n^{(1)}$ and $\nu_n^{(2)}$ can be obtained through perturbation theory. Let $|n\rangle_\pm$ be the eigenvector of $i\,\boldsymbol{\Omega}\,\boldsymbol{\nu}_{\oplus}$ associated to $\pm\nu_n$, and let $|n\rangle^\prime_\pm$ be the eigenvector of $i\,\boldsymbol{\Omega}\,\boldsymbol{\sigma}$ associated to $\pm\nu_n$. Also, we know from Williamson's theorem that $\boldsymbol{\sigma}=\boldsymbol{s}\,\boldsymbol{\nu}_{\oplus}\,\boldsymbol{s}^\dag$. We clearly have $i\,\boldsymbol{\Omega}\,\boldsymbol{\nu}^{(0)}_{\oplus}|n\rangle^{(0)}_\pm=\pm\nu_n^{(0)}\,|n\rangle^{(0)}_\pm$. Then, it is not difficult to show that $|n\rangle^{\prime(0)}_\pm=(\boldsymbol{s}^{(0)\dag})^{-1}\,|n\rangle^{(0)}_\pm=-\boldsymbol{\Omega}\,\boldsymbol{s}^{(0)}\,\boldsymbol{\Omega}\,|n\rangle^{(0)}_\pm$ and that ${}^{(0)}_\pm\langle n|^\prime={}^{(0)}_\pm\langle n|\boldsymbol{s}^{(0)\dag}$. Therefore, the general expression for the perturbative contributions to the symplectic eigenvalues is
\begin{align}\label{perturbative:symplectic:eigenvalue:computation:expression:first}
\nu_n^{(1)}:=&{}^{(0)}_+\langle n|^{\prime} i\,\boldsymbol{\Omega}\,\boldsymbol{\sigma}^{(1)}|n\rangle^{\prime(0)}_+
=-{}^{(0)}_+\langle n|\boldsymbol{s}^{(0)\dag}\,i\,\boldsymbol{\Omega}\,\boldsymbol{\sigma}^{(1)}\,\boldsymbol{\Omega}\,\boldsymbol{s}^{(0)}\,\boldsymbol{\Omega}\,|n\rangle^{(0)}_+\nonumber\\
\nu_n^{(2)}:=&{}^{(0)}_+\langle n|^{\prime} i\,\boldsymbol{\Omega}\,\boldsymbol{\sigma}^{(2)}|n\rangle^{\prime(0)}_++\sum_{k\neq n}\frac{\left|{}^{(0)}_+\langle k|^{\prime}i\,\boldsymbol{\Omega}\,\boldsymbol{\sigma}^{(1)}|n\rangle^{\prime(0)}_+\right|^2}{\nu_n^{(0)}-\nu_k^{(0)}},
\end{align}
which can be finally recast as
\begin{align}\label{perturbative:symplectic:eigenvalue:computation:expression}
\nu_n^{(1)}:=&-\sum_{pq\in\text{S}}\,\Omega_{qq}\,\Omega_{pp}\,s^{(0)*}_{pn}\,s^{(0)}_{qn}\,\sigma^{(1)}_{pq}\nonumber\\
\nu_n^{(2)}:=&-\sum_{pq\in\text{S}}\,\Omega_{qq}\,\Omega_{pp}\,s^{(0)*}_{pn}\,s^{(0)}_{qn}\,\sigma^{(2)}_{pq}+\sum_{k\neq n}\frac{\left|\sum_{pq\in\text{S}}\,\Omega_{qq}\,\Omega_{pp}\,s^{(0)*}_{pk}\,s^{(0)}_{qn}\,\sigma^{(1)}_{pq}\right|^2}{\nu_n^{(0)}-\nu_k^{(0)}}.
\end{align}

\section{Expressions for thermodynamical quantities in the perturbative regime}\label{appendix:one}
We are now in the position to compute the expressions for the thermodynamical quantities used in this work in the perturbative regime that we have adopted.

\subsection{Perturbative regime: energy}\label{appendix:one:perturbative:regime:energy}
Recall that the energy of the modes, when the state has zero first moments $\boldsymbol{d}$, reads \eqref{energy:expression:covariance:matrix}. Therefore, the energy of our system S reads
\begin{align}\label{energy:expression:covariance:matrix:appendix}
\Delta E_\textrm{S}=\frac{1}{2}\sum_{k\in \text{S}}\,\hbar\,\omega_k\,\left[\sigma_{\textrm{S},kk}(\lambda)-\sigma_{\textrm{S},kk}(0)\right].
\end{align}
It is easy to see that we can obtain \eqref{energy:expression:covariance:matrix:appendix} as
\begin{align}\label{energy:expression:covariance:matrix:appendix:new}
\Delta E_\textrm{S}=\frac{\hbar}{4}\textrm{Tr}\,\left(\boldsymbol{\omega}\,\boldsymbol{\sigma}_{\text{S}}(\lambda)-\boldsymbol{\omega}\,\boldsymbol{\sigma}_{\text{S}}(0)\right),
\end{align}
where we recall that $\boldsymbol{\omega}=\text{diag}(\omega_1,\ldots,\omega_n,\ldots,\omega_1,\ldots,\omega_n,\ldots)$ for the system S. 
The energy $E_\textrm{S}$ has the general perturbative  expansion $E_\textrm{S}=E_\textrm{S}^{(0)}+E_\textrm{S}^{(1)}\,h+ E_\textrm{S}^{(2)}\,h^2$ to second order in $h$.

When it is the case, as in this work, that $\Delta E_\textrm{S}^{(0)}=0$ we would have 
\begin{align}\label{energy:expression:covariance:matrix:appendix:perturbative:new}
\Delta E_\textrm{S}=\frac{\hbar}{4}\textrm{Tr}\,\left(\boldsymbol{\omega}\,\boldsymbol{\sigma}_{\text{S}}^{(1)}(\lambda)\,h+\boldsymbol{\omega}\,\boldsymbol{\sigma}_{\text{S}}^{(2)}(\lambda)\,h^2\right)=\frac{1}{2}\sum_{k\in \text{S}}\,\hbar\,\omega_k\,\left[\sigma^{(1)}_{\textrm{S},kk}(\lambda)\,h+\sigma^{(2)}_{\textrm{S},kk}(\lambda)\,h^2\right].
\end{align}

\subsection{Perturbative regime: entropy}\label{appendix:one:perturbative:regime:entropy}
We can proceed and compute the lowest order contribution to the change of entropy $\Delta S_\textrm{S}$ of the system S. We start by the definition of entropy \eqref{von:neumann:entropy} for Gaussian states
\begin{align}
S_\textrm{S}=&\sum_{k\in S}\left[\frac{\nu_{\text{S},k}+1}{2}\,\ln\left(\frac{\nu_{\text{S},k}+1}{2}\right)-\frac{\nu_{\text{S},k}-1}{2}\,\ln\left(\frac{\nu_{\text{S},k}-1}{2}\right)\right]
\end{align}
where $\nu_k$ are the symplectic eigenvalues of the state $\boldsymbol{\sigma}_\textrm{S}(\lambda)$. Notice the subscript S and note that the $\nu_k$ might depend on $\lambda$. We need to compute the symplectic eigenvalues \textit{not} of the full state but of the reduced state $\boldsymbol{\sigma}_\textrm{S}(\lambda)$ of the system. Therefore, it is easy to check that we have 
\begin{align}\label{entropy:change:first:order:appendix}
\Delta S_\textrm{S}=&\frac{1}{2}\,\sum_{n\in\text{S}}\,\left[\nu_{\text{S},n}^{(1)}\,\ln\left(\frac{\nu_{\text{S},n}^{(0)}+1}{\nu_{\text{S},n}^{(0)}-1}\right)\,h+\nu_{\text{S},n}^{(2)}\,\ln\left(\frac{\nu_{\text{S},n}^{(0)}+1}{\nu_{\text{S},n}^{(0)}-1}\right)\,h^2-\frac{\left(\nu_{\text{S},n}^{(1)}\right)^2}{\nu_{\text{S},n}^{(0)2}-1}\,h^2\right]\nonumber\\
=&\frac{1}{2}\,\sum_{n\in\text{S}}\,\left[\frac{\hbar\,\omega_n}{k_\text{B}\,T^{(0)}_n(\lambda)}\,\nu_{\text{S},n}^{(1)}\,h+\frac{\hbar\,\omega_n}{k_\text{B}\,T^{(0)}_n(\lambda)}\,\nu_{\text{S},n}^{(2)}\,h^2-\sinh^2\left(\frac{1}{2}\,\frac{\hbar\,\omega_n}{k_\text{B}\,T^{(0)}_n(\lambda)}\right)\,\left(\nu_{\text{S},n}^{(1)}\right)^2\,h^2\right]
\end{align}
for mixed states. Here we have used the fact that $\nu_{\text{S},n}^{(0)}=\coth\bigl(\frac{\hbar\,\omega_n}{2\,k_\text{B}\,T^{(0)}_n(\lambda)}\bigr)$ and we have inverted this expression.
The parameter $T^{(0)}_n(\lambda)$ is the lowest order contribution to the local temperature of mode $n$ in the reduced state of the system S \textit{after} the transformation $\boldsymbol{S}(\lambda)$ has applied.

Note that the expressions above are valid as long as every term in the perturbative series is small. This puts a limit on the values of the temperature that can be considered (compared to the small perturbative parameter $h$). For example, $\frac{\hbar\,\omega_n}{k_\text{B}\,T^{(0)}_n(\lambda)}\,\nu_{\text{S},n}^{(1)}\,h\ll1$ and so on. This tells us also that the limit for pure states needs to be addressed with care \cite{Safranek:Lee:2015}.

\subsection{Perturbative regime: efficiency}\label{appendix:one:perturbative:regime:efficiency}
We can now move on to the quantity of interest to this work. 
We combine all the elements of this section to provide the final expression for the efficiency $\eta$ in the perturbative regime when all changes in \eqref{efficiency:two}, except for the change in entropy, occur to first order in $h$. We find
\begin{align}\label{perturbative:efficiency:full}
\frac{1}{\eta}=1+\frac{\sum_{k\in E}k\,(\Delta N^{(1)}_k+\Delta N^{(2)}_k\,h)}{\sum_{k\in S}k\,(\Delta N^{(1)}_k+\Delta N^{(2)}_k\,h)},
\end{align}
where an explicit form of the quantities has been given above. We anticipate that there are many different and interesting behaviors of the expression \eqref{perturbative:efficiency:full} which depend dramatically on the nature of the initial state. We will study them in detail in the next section. 

We anticipate that we are interested in the general situation where $\Delta E_{\text{S}}^{(0)}=0$. We then see two possible scenarios that can occur. We list them here, while we study them in the next section.
\begin{itemize}
	\item[i)] The contributions $\Delta N^{(1)}_k$ vanish both in the numerators and denominator, which would give us an efficiency of the form
	\begin{align}\label{perturbative:efficiency:case:i}
		\frac{1}{\eta}&=1-\Gamma(t)+\mathcal{O}(h),
	\end{align}
where $\Gamma(t)$ is a time dependent function independent of $h$.
	\item[ii)] The term $\Delta N^{(1)}_k$ vanishes in the numerator but not in the denominator. Therefore, expression \eqref{efficiency:two} reads
\begin{align}\label{perturbative:efficiency:case:ii}
\frac{1}{\eta}&=1-\Gamma^{\prime}(t)\,h+\mathcal{O}(h^2),
\end{align}
where $\Gamma^{\prime}(t)$ is a time dependent function independent of $h$.
\end{itemize}
Below we will study these cases in more detail. 

\section{Perturbative emergence of fluctuations}\label{appen:peturbative mergence of fluctuations}
In order to determine the role of work fluctuations in our cavity we need a well defined physical model that allows us to obtain analytical expressions concerning the work distribution of the cavity. To this end we adhere to the following general assumptions on the systems Hamiltonian and its corresponding dynamics. 
\begin{itemize}
	\item[i)] The Hamiltonian is linear (i.e., quadratic in the creation and annihilation operators). This property preserves the Gaussian character of the state.
	\item[ii)] We focus on time evolution, although our work can be extended in a straightforward way to any quadratic transformation.
		\item[iii)] The Hamiltonian is cyclic. This means that $\hat{H}(0)=\hat{H}(T)$, while $\hat{H}(t)$ can take any quadratic form.
\end{itemize}
As discussed in section \ref{subsection:peturbative approach}, the disturbance of the confined quantum field may be described as a perturbation of its Hamiltonian
\begin{align}
	H^{(0)}\longrightarrow H^{(0)}+H^{(1)}h+H^{(2)}h^2+\mathcal{O}(h^3),
\end{align}
where $h$ is a small dimensionless parameter. The perturbation of the cavity Hamiltonian leading to a perturbation of the systems eigenstates and energy eigenvalues  in terms of the same parameter $h$ 
\begin{align}
\ket{n}^{(0)}&\longrightarrow\ket{n}^{(0)}+\ket{n}^{(1)}h+\ket{n}^{(2)}h^2+\mathcal{O}(h^3)\nonumber\\
E_n^{(0)}&\longrightarrow E_n^{(0)}+E_n^{(1)}h+E_n^{(2)}h^2+\mathcal{O}(h^3).
\end{align}
It is also important to see how this perturbation of the system's Hamiltonian affects the system's evolution, in particular from the definition of the unitary operator 
\begin{align}
 \hat{U}(t):=&\,\overset{\leftarrow}{\mathcal{T}}\exp\left[-\frac{i}{\hbar}\,\int_0^t\,dt'\,\hat{H}(t')\right]\nonumber\\ 
 =&\,\overset{\leftarrow}{\mathcal{T}}\exp\left[-\frac{i}{\hbar}\,\int_0^t\,dt'\,\left(H^{(0)}+H^{(1)}(t')h+H^{(2)}(t')h^2+\mathcal{O}(h^3)\right)\right],
 \end{align}
which we will use throughout this section.
 \subsection{Perturbed characteristic function}
 As described in the main text the characteristic function characterises the forward probability work distribution for the cavity after undergoing an evolution described by a perturbation of its Hamiltonian. We wish to expand the characteristic function at the lowest relevant perturbative orders in the perturbation parameter $h$ to determine when the work fluctuations start to play a role in our setup. The characteristic function is defined as
 \begin{align}
 \chi(u):=\sum_{m_q n_p}\,p_{m|n}\,p_n\,e^{-i\,(E^{f}_{m_q}-E^{i}_{n_p})\,u}.
 \end{align}
 The only two parameters which are functions of the perturbed Hamiltonian are the conditional probability distribution $p_{m|n}$ and the final energy eigenvalue $E^{f}_m$. We therefore expand the characteristic function up to second order in $h$, obtaining
 \begin{align}
 	\chi(u)\approx&\sum_{m_q,n_p}p_n\left(p^{(0)}_{m|n}+p^{(1)}_{m|n}h+p^{(2)}_{m|n}h^2\right)e^{-i\left(E_{m_q}^{f(0)}+E_{m_q}^{f(1)}h+E_{m_q}^{f(2)}h^2-E_{n_p}^{i}\right)u}\nonumber\\ 
 		\chi(u)\approx&\sum_{m_q,n_p}p_n e^{-i\left(E_{m_q}^{f(0)}-E_{n_p}^i\right)u}\left(p^{(0)}_{m|n}+p^{(1)}_{m|n}h+p^{(2)}_{m|n}h^2\right)e^{-i\left(E_{m_q}^{f(1)}h+E_{m_q}^{f(2)}h^2\right)u}.
 \end{align}
Noting that the definition of the conditional probability distribution is $p_{m|n}:=|\bra{m_q}U(t)\ket{n_p}|^2$, the zeroth contribution of this term being $p^{(0)}_{m|n}=|^{(0)}\bra{m_q}\ket{n_p}^{(0)}|^2=\delta_{m_q,n_p}$, and later in section \ref{section:peturbed conditional prob} we also show that the conditional probability distribution has no first order contribution, $p_{m|n}^{(1)}=0$. This simplifies the above characteristic function to the expression
 \begin{align}
\chi(u)\approx&\sum_{m_q,n_p}p_n e^{-i\left(E_{m_q}^{f(0)}-E_{n_p}^i\right)u}\left(\delta_{m_q,n_p}+p^{(2)}_{m|n}h^2\right)e^{-i\left(E_{m_q}^{f(1)}h+E_{m_q}^{f(2)}h^2\right)u}.
\end{align}
Making use of the identity $e^{\psi+\phi}=e^{\psi}+\left(e^\phi-1\right)+\mathcal{O}(\psi\phi)$, allows us to write the characteristic function to second order as 
\begin{align}\label{equ:characteristic fucntion to 2nd order}
\chi(u)\approx1+\sum_{n_p}p_n\left(e^{-iE_{n_p}^{f(1)}hu}-1\right)+\sum_{m_q,n_p}p_n  p_{m|n}^{(2)}h^2e^{-i\left(E_{m_q}^{f(0)}-E_{n_p}^i\right)u}+\sum_{n_p} p_n \left(e^{-iE_{n_p}^{f(2)}h^2u}-1\right).
\end{align}
In order to extract meaningful information from the characteristic function we can calculate its moments using equation \eqref{moment}, where $n=1$ gives the mean of the distribution and $n=2$ gives the distributions variance. These two quantities are therefore have perturbative expansion of the form
\begin{align}
	G\left[\psi\right]=&\sum_{n_p} p_n E_{n_p}^{f(1)}h+\sum_{n_p} p_n E_{n_p}^{f(2)}h^2+\sum_{m_q,n_p}p_n p_{m|n}^{(2)}\left(E_{m_q}^{f(0)}-E_{n_p}^i\right)h^2\label{equ:1moment}\\
G\left[\psi^2\right]=&\sum_{n_p}p_n\left(E_{n_p}^{f(1)}\right)^2h^2+\sum_{m_q,n_p}p_np_{m|n}^{(2)}\left(E_{m_q}^{f(0)}-E_{n_p}^i\right)^2h^2.\label{equ:2moment}
\end{align}
All that remain is to calculate the quantities $p_{m|n}^{(2)},E_n^{f(1)},E_n^{f(2)}$ explicitly in terms of the Bogoliubov coefficients. 
\subsection{Perturbed Hamiltonians} 
In order to calculate the quantities $p_{m|n}^{(2)},E_n^{f(1)},E_n^{f(2)}$ explicitly in terms of the Bogoliubov coefficients describing the cavity dynamics we need to relate the perturbed Hamiltonian matrix to the elements of the symplectic matrix that represent the Unitary evolution.

We start by employing the definition of the symplectic matrix in terms of the Hamiltonian matrix \cite{Bruschi:Lee:2013}, which in general reads
\begin{align}\label{symplectic:representation:Hamiltonian}
\boldsymbol{S}(t)=&\overleftarrow{T}\exp(\boldsymbol{\Omega}\int^{t}_{0}dt'\boldsymbol{{H}}(t')).
\end{align} 
A perturbative expansion of the Hamiltonian and of the symplectic matrix allows us to find
\begin{align}
\boldsymbol{S}^{(0)}+\boldsymbol{S}^{(1)}\,h+\boldsymbol{S}^{(2)}\,h^2=&\overleftarrow{T}\exp(\boldsymbol{\Omega}\,\int_{0}^{t}dt'(\boldsymbol{{H}}^{(0)}+\boldsymbol{{H}}^{(1)}h+\boldsymbol{{H}}^{(2)}h^2))\nonumber\\
=&\boldsymbol{S}^{(0)}\overleftarrow{T}\exp(\boldsymbol{\Omega }\,\int_{0}^{t}dt'\boldsymbol{S}^{(0)\dagger}(\boldsymbol{{H}}^{(1)}h+\boldsymbol{{H}}^{(2)}h^2)\boldsymbol{S}^{(0)})\nonumber\\
=&\boldsymbol{S}^{(0)}\Bigg{[}\overleftarrow{T}\exp(\boldsymbol{\Omega}\,\int_{0}^{t}dt'\boldsymbol{S}^{(0)\dagger}\boldsymbol{{H}}^{(1)}\boldsymbol{S}^{(0)}h)\overleftarrow{T}\exp(\boldsymbol{\Omega}\,\int_{0}^{t}dt'\boldsymbol{S}^{(0)\dagger}\boldsymbol{{H}}^{(2)}\boldsymbol{S}^{(0)}h^2)\Bigg{]}\nonumber\\
=&\boldsymbol{S}^{(0)}\left(\mathds{1}+\boldsymbol{\Omega }\int_{0}^{t}dt'\boldsymbol{S}^{(0)\dagger}\boldsymbol{{H}}^{(1)}\boldsymbol{S}^{(0)}h+\boldsymbol{\Omega }\,\int_{0}^{t}dt'\boldsymbol{S}^{(0)\dagger}\boldsymbol{{H}}^{(1)}\boldsymbol{S}^{(0)}\boldsymbol{\Omega }\int_{0}^{t'}dt''\boldsymbol{S}^{(0)\dagger}\boldsymbol{{H}}^{(1)}\boldsymbol{S}^{(0)}h^2\right.\nonumber\\
&+\left.\boldsymbol{\Omega }\int_{0}^{t}dt'\boldsymbol{S}^{(0)\dagger}\boldsymbol{{H}}^{(2)}\boldsymbol{S}^{(0)}h^2\right),
\end{align} 
where we have been able to separate the exponentials proportional to $h$ and $h^2$ since their commutator gives corrections to $\mathcal{O}(h^3)$, which we discard. Note that $[\boldsymbol{\Omega},\boldsymbol{S}^{(0)}]=0$.
This expansion allows us to immediately identify each contribution and we obtain 
\begin{align}
\label{1stHam}\boldsymbol{S}^{(1)}=&\boldsymbol{\Omega }\,\boldsymbol{S}^{(0)}\,\int_{0}^{t}dt'\boldsymbol{S}^{(0)\dagger}\boldsymbol{{H}}^{(1)}\boldsymbol{S}^{(0)},\\
\label{2ndHam}\boldsymbol{S}^{(2)}=&\boldsymbol{S}^{(0)}\Big[\boldsymbol{\Omega}\,\int_{0}^{t}dt'\boldsymbol{S}^{(0)\dagger}\boldsymbol{{H}}^{(1)}\boldsymbol{S}^{(0)}\boldsymbol{\Omega }\int_{0}^{t'}dt''\boldsymbol{S}^{(0)\dagger}\boldsymbol{{H}}^{(1)}\boldsymbol{S}^{(0)}+\boldsymbol{\Omega }\int_{0}^{t}dt'\boldsymbol{S}^{(0)\dagger}\boldsymbol{H}^{(2)}\boldsymbol{S}^{(0)}\Big].
\end{align}
Using the identities $\boldsymbol{S}^{(0)\dagger}\boldsymbol{S}^{(0)}=\mathds{1}$ and $\boldsymbol{\Omega}^2=-\mathds{1}$, we can employ some algebra and show that \eqref{1stHam} gives
\begin{align}\label{1stHamb}
\boldsymbol{{H}}^{(1)}=&\boldsymbol{\Omega }(\boldsymbol{S}^{(0)}\dot{\boldsymbol{S}}^{(0)\dagger}\boldsymbol{S}^{(1)}\boldsymbol{S}^{(0)\dagger}+\dot{\boldsymbol{S}}^{(1)}\boldsymbol{S}^{(0)\dagger}).
\end{align}
To obtain this we have used the fact that
\begin{align}\label{H1sub}
\boldsymbol{\Omega }\int_{0}^{t}dt'\boldsymbol{S}^{(0)\dagger}\boldsymbol{{H}}^{(1)}\boldsymbol{S}^{(0)}=\boldsymbol{S}^{(0)\dagger}\boldsymbol{S}^{(1)}\,\,\,\Leftrightarrow\,\,\,
\boldsymbol{S}^{(0)\dagger}\boldsymbol{{H}}^{(1)}\boldsymbol{S}^{(0)}=-\frac{d}{dt}(\boldsymbol{\Omega }\boldsymbol{S}^{(0)\dagger}\boldsymbol{S}^{(1)}),
\end{align}
rearranging for $\boldsymbol{{H}}^{(1)}$.

The second order Hamiltonian can be derived in a similar fashion from equation \eqref{2ndHam} and after some algebra we find
\begin{align}
\boldsymbol{{H}}^{(2)}=&-\boldsymbol{\Omega }\,\left[\boldsymbol{S}^{(0)}\dot{\boldsymbol{S}}^{(0)\dagger}\boldsymbol{S}^{(2)}\boldsymbol{S}^{(0)\dagger}+\dot{\boldsymbol{S}}^{(2)}\boldsymbol{S}^{(0)\dagger}+\boldsymbol{S}^{(0)}\dot{\boldsymbol{S}}^{(0)\dagger}\boldsymbol{S}^{(1)}\boldsymbol{S}^{(0)\dagger}\boldsymbol{S}^{(1)}\boldsymbol{S}^{(0)\dagger}+\dot{\boldsymbol{S}}^{(1)}\boldsymbol{S}^{(0)\dagger}\boldsymbol{S}^{(1)}\boldsymbol{S}^{(0)\dagger}\right].
\end{align}
The expressions for arbitrary quadratic contributions $\boldsymbol{H}^{(1)}$ and $\boldsymbol{H}^{(2)}$ are obtained from the general expressions for a quadratic Hamiltonian, that is, 
\begin{align}\label{A:quadham}
\hat{H}^{(1)}=&\sum_{kl}\,\left[M^{(1)}_{kl}\,\hat{a}^{\dag}_k\,\hat{a}^{\dag}_l+N^{(1)}_{kl}\,\hat{a}^{\dag}_k\,\hat{a}_l+P^{(1)}_{kl}\,\hat{a}_k\,\hat{a}^{\dag}_l+M^{(1)*}_{kl}\,\hat{a}_k\,\hat{a}_l\right],\nonumber\\
\hat{H}^{(2)}=&\sum_{kl}\,\left[M^{(2)}_{kl}\,\hat{a}^{\dag}_k\,\hat{a}^{\dag}_l+N^{(2)}_{kl}\,\hat{a}^{\dag}_k\,\hat{a}_l+P^{(2)}_{kl}\,\hat{a}_k\,\hat{a}^{\dag}_l+M^{(2)*}_{kl}\,\hat{a}_k\,\hat{a}_l\right].
\end{align}  
Employing the expressions obtained above we have the following relations for the first order terms 
\begin{align}\label{block_matrix}
N^{(1)}_{kl}&=\,-i\,\left[{\alpha}^{(0)}_{kk}\,\dot{{\alpha}}^{(0)*}_{kk}{\alpha}^{(1)}_{kl}\,{\alpha}^{(0)*}_{ll}\,+\,\dot{{\alpha}}^{(1)}_{kl}\,{\alpha}^{(0)*}_{ll}\right],\nonumber\\ 
M^{(1)}_{kl}&=\,-i\,\left[\,{\alpha}^{(0)}_{kk}\,\dot{{\alpha}}^{(0)*}_{kk}\,\,{\beta}^{(1)}_{kl}\,\,{\alpha}^{(0)}_{ll}\,+\,\dot{{\beta}}^{(1)}_{kl}\,\,{\alpha}^{(0)}_{ll}\right],\nonumber\\ 
P^{(1)}_{kl}&=\,i\,\left[\,{\alpha}^{(0)*}_{kk}\,{\dot{\alpha}}^{(0)}_{kk}\,\alpha^{(1)\dagger}_{kl}\,{\alpha}^{(0)}_{ll}\,+\,\dot{{\alpha}}^{(1)\dagger}_{kl}\,{\alpha}^{(0)}_{ll}\right],  
\end{align} 
and the second order ones 
\begin{align}\label{block_matrix2}
N^{(2)}_{kl}&=\,i\,\left[{\alpha}^{(0)}_{kk}\,\dot{{\alpha}}^{(0)*}_{kk}{\alpha}^{(2)}_{kl}\,{\alpha}^{(0)*}_{ll}\,+\,\dot{{\alpha}}^{(2)}_{kl}\,{\alpha}^{(0)*}_{ll}+{\alpha}^{(0)}_{kk}\,\dot{{\alpha}}^{(0)*}_{kk}\,\sum_{j}{\alpha}^{(1)}_{kj}\,{\alpha}^{(0)*}_{jj}\,{\alpha}^{(1)}_{jl}\,{\alpha}^{(0)*}_{ll}+\sum_j\dot{{\alpha}}^{(1)}_{kj}\,{\alpha}^{(0)*}_{jj}\,{\alpha}^{(1)}_{jl}\,{\alpha}^{(0)*}_{ll}\right],\nonumber\\ 
M^{(2)}_{kl}&=\,-i\,\left[\,{\alpha}^{(0)}_{kk}\,\dot{{\alpha}}^{(0)*}_{kk}\,\,{\beta}^{(2)}_{kl}\,\,{\alpha}^{(0)}_{ll}\,+\,\dot{{\beta}}^{(2)}_{kl}\,\,{\alpha}^{(0)}_{ll}-{\alpha}^{(0)}_{kk}\,\dot{{\alpha}}^{(0)*}_{kk}\sum_j\,{\beta}^{(1)}_{kj}\,{\alpha}^{(0)*}_{jj}\,{\beta}^{(1)}_{jl}\,{\alpha}_{ll}^{(0)*}-\sum_j\dot{{\beta}}^{(1)}_{kj}\,{\alpha}_{jj}^{(0)*}\,{\beta}_{jl}^{(1)}\,{\alpha}_{ll}^{(0)*}\right],\nonumber\\ 
P^{(2)}_{kl}&=\,i\,\left[\,{\alpha}^{(0)*}_{kk}\,{\dot{\alpha}}^{(0)}_{kk}\,\alpha^{(2)\dagger}_{kl}\,{\alpha}^{(0)}_{ll}\,+\,\dot{{\alpha}}^{(2)\dagger}_{kl}\,{\alpha}^{(0)}_{ll}-{\alpha}^{(0)*}_{kk}\,\dot{{\alpha}}^{(0)}_{kk}\sum_j\,{\alpha}^{(1)\dagger}_{kj}\,{\alpha}^{(0)}_{jj}\,{\alpha}^{(1)\dagger}_{jl}\,{\alpha}^{(0)}_{ll}-\sum_j\dot{{\alpha}}^{(1)\dagger}_{kj}\,{\alpha}^{(0)}_{jj}\,{\alpha}^{(1)\dagger}_{jl}\,{\alpha}^{(0)}_{ll}\right].  
\end{align}
\subsection{Perturbed energy eigenvalue}
The definition of the energy eigenvalue for is
\begin{align}
\hat{H}(t) |n_p \rangle_i=E_{n_p}(t) |n_p \rangle_i 
\end{align}
perturbing the variables such that $\hat{H}=\hat{H}^{(0)}+\hat{H}^{(1)}\,h+\hat{H}^{(2)}\,h^2+\mathcal{O}(h^3)$, $E_{n_p}=E_{n_p}^{(0)}+E_{n_p}^{(1)}h+E_{n_p}^{(2)}h^2+\mathcal{O}(h^3)$ and $ |n_p \rangle_i =|n_p \rangle^{(0)}+|n_p \rangle^{(1)}\,h+|n_p \rangle^{(2)}\,h^2+\mathcal{O}(h^3) $ and operating through by ${}^{(0)}\langle n_p| $ we get 
\begin{align}
E_{n_p}^{(1)}(t)&={}^{(0)}\langle n_p|\hat{H}^{(1)}|n_p\rangle^{(0)}.
\end{align}
Employing from \eqref{A:quadham} the definition of the first order Hamiltonian we get for the first order energy eigenvalue
\begin{align}\label{equ:first order energy eigenvalue}
E^{(1)}_{n_p}=n_p\,(N^{(1)}_{pp}+P^{(1)}_{pp})+\sum_k\,P^{(1)}_{kk}.
\end{align}
Again, taking the definition of the energy eigenvalue but this time expanding to second order we find
{\small
\begin{align}
(\hat{H}^{(0)}+\hat{H}^{(1)}\,h+\hat{H}^{(2)}\,h^2)\,(|n_p \rangle^{(0)}+|n_p \rangle^{(1)}\,h+|n_p \rangle^{(2)}\,h^2)=&
(E_{n_p}^{(0)}+E_{n_p}^{(1)}h+E_{n_p}^{(2)}h^2)\,(|n_p \rangle^{(0)}+|n_p \rangle^{(1)}\,h+|n_p \rangle^{(2)}\,h^2)\nonumber\\ 
H^{(0)}\,|n_p\rangle^{(2)}+H^{(1)}\,|n_p\rangle^{(1)}+H^{(2)}\,|n_p\rangle^{(0)}=&E_{n_p}^{(0)}\,|n_p\rangle^{(2)}+E_{n_p}^{(1)}\,|n_p\rangle^{(1)}+E_{n_p}^{(2)}\,|n_p\rangle^{(0)}.
\end{align}
}
Projecting the above equation onto the vector ${}^{(0)}\langle{n_p}|$, we find 
\begin{align}\label{C5}
E_{n_p}^{(2)}={}^{(0)}\langle n_p |\,H^{(1)}\,|n_p\rangle^{(1)}+{}^{(0)}\langle n_p |\,H^{(2)}\,|n_p\rangle^{(0)}. 
\end{align}
To derive the contribution $|n_p\rangle^{(1)}$, standard perturbation theory \cite{landau2013quantum} uses the resolution of the identity in a single subspace $\mathds{1}=\sum_i |i\rangle\langle i|$ however, since our cavity consists of infinite modes, the resolution of the identity must occur over every subspace, and therefore we have 
\begin{align}\label{C6}
|n_p\rangle^{(1)}=\sum_{\{\}\neq n_p}\frac{{}^{(0)}\langle ...m_r ... |\,H^{(1)}\,|n_p\rangle^{(0)}}{{E_{n_p}^{(0)}-E_{m_r}^{(0)}}}\,|...m_r ...\rangle^{(0)},
\end{align} 
where $\{\}$ denotes an infinite collection of modes of the form $|...m_r...\rangle$ and $\{\}\neq n_p$ means that $\sum_r m_r\neq n_p$.

Substituting into $H^{(1)}$ gives us 
{\small
\begin{align}
|n_p\rangle^{(1)}=&\sum_{\{\}\neq n_p}\frac{|...m_r ...\rangle^{(0)}}{{E_{n_p}^{(0)}-E_{m_r}^{(0)}}}\,\sum_{kl}\,\Big{\{}\,M_{kl}^{(1)}\,{}^{(0)}\langle ...m_r ... |\,a_k^{\dagger}\,a_l^{\dagger}\,|n_p\rangle^{(0)}+M_{kl}^{(1)*}\,{}^{(0)}\langle ...m_r ... |\,a_k\,a_l\,|n_p\rangle^{(0)}\nonumber\\ 
&+N_{kl}^{(1)}\,{}^{(0)}\langle ...m_r ... |\,a_k^{\dagger}\,a_l\,|n_p\rangle^{(0)}+P_{kl}^{(1)}\,{}^{(0)}\langle ...m_r ... |\,a_k\,a_l^{\dagger}\,|n_p\rangle^{(0)}\,\Big{\}}\nonumber\\ 
=&\sum_{\{\}\neq n_p}\frac{|...m_r ...\rangle^{(0)}}{{E_{n_p}^{(0)}-E_{m_r}^{(0)}}}\,\sum_{kl}\,\Big{\{}\,M_{pp}^{(1)}\,\sqrt{n_p+1}\sqrt{n_p+2}\delta_{kp}\delta_{lp}{}^{(0)}\langle ...m_r ... |n_p+2\rangle^{(0)}\nonumber\\
&+\sum_{k\neq p}M_{kp}^{(1)}\,\sqrt{n_p+1}\delta_{lp}\,{}^{(0)}\langle ...m_r ... |n_p+1,1_k\rangle^{(0)}+\sum_{l\neq p}M_{pl}^{(1)}\,\sqrt{n_p+1}\delta_{kp}\,{}^{(0)}\langle ...m_r ... |n_p+1,1_l\rangle^{(0)}\nonumber\\
&+\sum_{k,l\neq p}M_{kl}^{(1)}\,{}^{(0)}\langle ...m_r ... |n_p,1_k,1_l\rangle^{(0)}(1-\delta_{kp})(1-\delta_{kl})+M_{pp}^{(1)*}\,\sqrt{n_p}\,\sqrt{n_p-1}\delta_{kp}\delta_{lp}{}^{(0)}\langle ...m_r ... |n_p-2\rangle^{(0)}\nonumber\\ 
&+N_{kp}^{(1)}\,\sqrt{n_p}\delta_{lp}{}^{(0)}\langle ...m_r ... |n_p-1,1_k\rangle^{(0)}+N_{pp}^{(1)}\,n_p\delta_{kp}\delta_{lp}\,{}^{(0)}\langle ...m_r ... |n_p\rangle^{(0)}\nonumber\\ 
&+P_{kl}^{(1)}\,\delta_{kl}{}^{(0)}\langle ...m_r ... |n_p\rangle^{(0)}+P_{pl}^{(1)}\,\sqrt{n_p}\delta_{kp}{}^{(0)}\langle ...m_r ... |n_p-1,1_l\rangle^{(0)}+P_{pp}^{(1)}\,(n_p+1)\delta_{kp}\delta_{lp}\,{}^{(0)}\langle ...m_r ... |n_p\rangle^{(0)}\,\Big{\}}. 
\end{align}
}
Using the fact that $\{\}\neq n_p$, we can reduce the above equation to
\begin{align}\label{statepert}
|n_p\rangle^{(1)}=&\frac{\sqrt{n_p+1}\sqrt{n_p+2}}{-E_{2_p}^{(0)}}M^{(1)}_{pp}|n_p+2\rangle^{(0)}
+\sum_{k\neq p}\frac{\sqrt{n_p+1}}{-E^{(0)}_{1_p,1_k}}M_{kp}^{(1)}|n_p+1,1_k\rangle^{(0)}\nonumber\\
&+\sum_{l\neq p}\frac{\sqrt{n_p+1}}{-E^{(0)}_{1_p,1_l}}M_{pl}^{(1)}|n_p+1,1_l\rangle^{(0)}+\sum_{k,l\neq p}\frac{1}{-E^{(0)}_{1_k,1_l}}M^{(1)}_{kl}|n_p,1_k,1_l\rangle^{(0)}\nonumber\\
&+\frac{\sqrt{n_p}\sqrt{n_p-1}}{E_{2_p}^{(0)}}M^{(1)*}_{pp}|n_p-2\rangle^{(0)}+\sum_{k\neq p}\frac{\sqrt{n_p}}{E_{1_p}^{(0)}-E^{(0)}_{1_k}}N^{(1)}_{kp}|n_p-1,1_k\rangle^{(0)}\nonumber\\ 
&+\sum_{l\neq p}\frac{\sqrt{n_p}}{E^{(0)}_{1_p}-E^{(0)}_{1_l}}P^{(1)}_{pl}|n_p-1,1_l\rangle^{(0)}
\end{align}
which, when substituted into \eqref{C5}, gives us 
\begin{align}
E_{n_p}^{(2)}=&\frac{\sqrt{n_p+1}\sqrt{n_p+2}}{-E_{2_p}^{(0)}}M^{(1)}_{pp}{}^{(0)}\langle n_p| H^{(1)}|n_p+2\rangle^{(0)}
+\sum_{k\neq p}\frac{\sqrt{n_p+1}}{-E^{(0)}_{1_p,1_k}}M_{kp}^{(1)}{}^{(0)}\langle n_p| H^{(1)}|n_p+1,1_k\rangle^{(0)}\nonumber\\
&+\sum_{l\neq p}\frac{\sqrt{n_p+1}}{-E^{(0)}_{1_p,1_l}}M_{pl}^{(1)}{}^{(0)}\langle n_p| H^{(1)}|n_p+1,1_l\rangle^{(0)}+\sum_{k,l\neq p}\frac{1}{-E^{(0)}_{1_k,1_l}}M^{(1)}_{kl}{}^{(0)}\langle n_p| H^{(1)}|n_p,1_k,1_l\rangle^{(0)}\nonumber\\
&+\frac{\sqrt{n_p}\sqrt{n_p-1}}{E_{2_p}^{(0)}}M^{(1)*}_{pp}{}^{(0)}\langle n_p| H^{(1)}|n_p-2\rangle^{(0)}+\sum_{k\neq p}\frac{\sqrt{n_p}}{E_{1_p}^{(0)}-E^{(0)}_{1_k}}N^{(1)}_{kp}{}^{(0)}\langle n_p| H^{(1)}|n_p-1,1_k\rangle^{(0)}\nonumber\\ 
&+\sum_{l\neq p}\frac{\sqrt{n_p}}{E^{(0)}_{1_p}-E^{(0)}_{1_l}}P^{(1)}_{pl}{}^{(0)}\langle n_p| H^{(1)}|n_p-1,1_l\rangle^{(0)}+{}^{(0)}\langle n_p |\,H^{(2)}\,|n_p\rangle^{(0)}.
\end{align} 
Now substituting the quadratic Hamiltonians $H^{(1)},H^{(2)}$ gives us 
\begin{align}\label{equ:second order energy eigenvalue}
E_{n_p}^{(2)}=&-2\,\frac{2\,n_p+1}{E_{2_p}^{(0)}}|M^{(1)}_{pp}|^2
-4\,\sum_{k\neq p}\frac{n_p+1}{E^{(0)}_{1_p,1_k}}\,|M_{kp}^{(1)}|^2-2\,\sum_{k,l\neq p}\frac{1}{E^{(0)}_{1_k,1_l}}\,|M^{(1)}_{kl}|^2+\sum_{k\neq p}\frac{n_p}{E_{1_p}^{(0)}-E^{(0)}_{1_k}}\,\left|N_{pk}^{(1)}+P_{kp}^{(1)}\right|^2\nonumber\\
&+n_p\,(N^{(2)}_{pp}+P^{(2)}_{pp})+\sum_k\,P^{(2)}_{kk}.
\end{align} 
This expression is complicated but it still has a manageable form.
\subsection{Perturbed conditional probability}\label{section:peturbed conditional prob}
The definition of the conditional probability distribution is
\begin{align} 
p_{m|n}(t)=|{}_f\langle m_q | \hat{U}(t)|n_p\rangle_i|^2. 
\end{align}
We take the first order expansion of ${}_f\langle m_q |$, and the unitary operator $ \hat{U}(t)$ using the Dyson series \cite{dyson1949radiation}, to obtain
\begin{align} 
p_{m|n}(t)&=\left|\left(\langle m_q|^{(0)}+\langle m_q|^{(1)}h \right)\left(\mathds{1}-\frac{i}{\hbar}\int_{0}^{t}dt' \hat{H}^{(1)}(t')h\right)\left(|n_p \rangle^{(0)}\right)\right|^2+\mathcal{O}(h^2),\nonumber\\
&=\left|{}^{(0)}\langle m_q|n_p\rangle^{(0)}+{}^{(1)}\langle m_q|n_p\rangle^{(0)}h-\frac{i}{\hbar}\int_{0}^{t}dt' {}^{(0)}\langle m_q|\hat{H}^{(1)}(t')|n_p\rangle^{(0)}h\right|^2+\mathcal{O}(h^2)\nonumber\\
&=\left|\delta_{m_q,n_p}-\frac{i}{\hbar}\int_{0}^{t}dt' {}^{(0)}\langle m_q|\hat{H}^{(1)}(t')|n_p\rangle^{(0)}h\right|^2+\mathcal{O}(h^2),
\end{align}
where ${}^{(1)}\langle m_q|n_p\rangle^{(0)}=0$. Expanding the absolute square and taking the first order contribution 
\begin{align}\label{equ:p1=0}
p^{(1)}_{m|n}(t)=&  \delta_{m_q,n_p}\frac{i}{\hbar}\int_{0}^{t}dt' E^{(1)}_{n_p}(t')h- \delta_{m_q,n_p}\frac{i}{\hbar}\int_{0}^{t}dt' E^{(1)}_{n_p}(t')h,\nonumber\\ 
=& 0.
\end{align}
We continue by computing the second order contribution to the conditional probability $p_{m|n}$ for one mode.
Starting with the unitary operator to second order 
\begin{align}\label{U_2_b}
\hat{U}(t)&=\hat{U}_0\overleftarrow{T}\exp\left[{-\frac{i}{\hbar}\int_{0}^{t}dt'\big{(}\hat{U}_0^\dagger \hat{H}^{(1)}\,\hat{U}_0\,h+\hat{U}_0^\dagger\,\hat{H}^{(2)}\,\hat{U}_0h^2\big{)}}\right],
\end{align}
we expand the exponential in \eqref{U_2_b} by means of \cite{dyson1949radiation}, and recall that we are ignoring terms of $\mathcal{O}(h^3)$ to find
{\small
\begin{align}\label{main:formula:two:one:b;c3}
{}_\textrm{f}\langle m_q|\,\hat{U}(t)\,|n_p\rangle_\textrm{i}=&{}_\textrm{f}\langle m_q|\,\hat{U}^{(0)}(t)\,|n_p\rangle_\textrm{i}+\frac{i}{\hbar}\,{}_\textrm{f}\langle m_q|\int_0^tdt'\,\hat{U}^{(0)\,\dag}(t')\,\hat{H}^{(1)}(t')\,\hat{U}^{(0)}(t')\,\hat{U}^{(0)\dag}(t)\,|n_p\rangle_\textrm{i}\,h\nonumber\\ &+\frac{i}{\hbar}\,{}_\textrm{f}\langle m_q|\int_0^tdt'\,\hat{U}^{(0)\,\dag}(t')\,\hat{H}^{(2)}(t')\,\hat{U}^{(0)}(t')\,\hat{U}^{(0)\dag}(t)\,|n_p\rangle_\textrm{i}\,h^2\,\nonumber\\
&-\frac{1}{\hbar^2}\,{}_\textrm{f}\langle m_q|\int_0^t\int_0^{t'}dt'dt''\Big{[}\,\hat{U}^{(0)\,\dag}(t')\,\hat{H}^{(1)}(t')\,\hat{U}^{(0)}(t') \hat{U}^{(0)\,\dag}(t'')\,\hat{H}^{(1)}(t'')\,\hat{U}^{(0)}(t'')\,\Big{]}\hat{U}^{(0)\dag}(t)\,|n_p\rangle_\textrm{i}\,h^2.
\end{align}
}
We know that $\hat{U}^{(0)\dag}(t)\,\hat{a}_p\,\hat{U}^{(0)}(t)=\exp[-i\,\omega_p\,t]\,\hat{a}_p$ and that $|n_p\rangle^{(0)}:=\frac{\hat{a}_p^{\dag n}}{\sqrt{n!}}|0\rangle$. Expanding ${}_f\bra{m_q}$, equation \eqref{main:formula:two:one:b;c3} reduces to
{\small
\begin{align}\label{main:formula:two:one:b;c4a}
{}_\textrm{f}\langle m_q|\,\hat{U}(t)\,|n_p\rangle_\textrm{i}=&\left(\langle m_q|^{(0)}+\langle m_q|^{(1)}h +\bra{m_q}^{(2)}h^2\right)\,\hat{U}^{(0)}(t)\,|n_p \rangle^{(0)}\nonumber\\
&+\frac{i}{\hbar}\,\left(\langle m_q|^{(0)}+\langle m_q|^{(1)}h \right)\int_0^tdt'\,\hat{U}^{(0)\,\dag}(t')\,\hat{H}^{(1)}(t')\,\hat{U}^{(0)}(t')\,\hat{U}^{(0)\dag}(t)\,|n_p \rangle^{(0)}\,h\nonumber\\ &+\frac{i}{\hbar}\,{}^{(0)}\langle m_q|\int_0^tdt'\,\hat{U}^{(0)\,\dag}(t')\,\hat{H}^{(2)}(t')\,\hat{U}^{(0)}(t')\,\hat{U}^{(0)\dag}(t)\,|n_p \rangle^{(0)}\,h^2\,\nonumber\\
&-\frac{1}{\hbar^2}\,{}^{(0)}\langle m_q|\int_0^t\int_0^{t'}dt'dt''\Big{[}\,\hat{U}^{(0)\,\dag}(t')\,\hat{H}^{(1)}(t')\,\hat{U}^{(0)}(t') \hat{U}^{(0)\,\dag}(t'')\,\hat{H}^{(1)}(t'')\,\hat{U}^{(0)}(t'')\,\Big{]}\hat{U}^{(0)\dag}(t)\,|n_p\rangle^{(0)}\,h^2,\nonumber\\ 
=&{}^{(0)}\langle m_q|\,\hat{U}^{(0)}(t)\,|n_p\rangle^{(0)}+{}^{(1)}\langle m_q|\,\hat{U}^{(0)}(t)\,|n_p\rangle^{(0)}h+{}^{(2)}\langle m_q|\,\hat{U}^{(0)}(t)\,|n_p\rangle^{(0)}h^2\nonumber\\
&+\frac{i}{\hbar}\,{}^{(0)}\langle m_q|\int_0^tdt'\,\hat{U}^{(0)\,\dag}(t')\,\hat{H}^{(1)}(t')\,\hat{U}^{(0)}(t')\,\hat{U}^{(0)\dag}(t)\,|n_p \rangle^{(0)}\,h\nonumber\\ 
&+\frac{i}{\hbar}\,{}^{(1)}\langle m_q|\int_0^tdt'\,\hat{U}^{(0)\,\dag}(t')\,\hat{H}^{(1)}(t')\,\hat{U}^{(0)}(t')\,\hat{U}^{(0)\dag}(t)\,|n_p \rangle^{(0)}\,h^2\nonumber\\ 
&+\frac{i}{\hbar}\,{}^{(0)}\langle m_q|\int_0^tdt'\,\hat{U}^{(0)\,\dag}(t')\,\hat{H}^{(2)}(t')\,\hat{U}^{(0)}(t')\,\hat{U}^{(0)\dag}(t)\,|n_p \rangle^{(0)}\,h^2\,\nonumber\\
&-\frac{1}{\hbar^2}\,{}^{(0)}\langle m_q|\int_0^t\int_0^{t'}dt'dt''\Big{[}\,\hat{U}^{(0)\,\dag}(t')\,\hat{H}^{(1)}(t')\,\hat{U}^{(0)}(t') \hat{U}^{(0)\,\dag}(t'')\,\hat{H}^{(1)}(t'')\,\hat{U}^{(0)}(t'')\,\Big{]}\hat{U}^{(0)\dag}(t)\,|n_p\rangle^{(0)}\,h^2,
\end{align}
}
up to third order. Adding the phases where and noting that ${}^{(1)}\langle m_q|n_p\rangle^{(0)}={}^{(2)}\langle m_q|n_p\rangle^{(0)}=0$,  we can write
\begin{align}\label{equ:unitary thing}
{}_\textrm{f}\langle m_q|\,\hat{U}(t)\,|n_p\rangle_\textrm{i}=&e^{-i\omega_pn_p t}\delta_{m_q n_p}\,\delta_{p q}\nonumber\\
&+\frac{i}{\hbar}\,\int_0^tdt'\,e^{-i\omega_qm_q t'}\,e^{i\omega_pn_p t}\,e^{-i\omega_pn_p t'}\,{}^{(0)}\bra{m_q}\hat{H}^{(1)}(t')\,|n_p \rangle^{(0)}\,h\nonumber\\ 
&+\frac{i}{\hbar}\,\int_0^tdt'e^{i\omega_pn_p t}\,e^{-i\omega_pn_p t'}\,{}^{(1)}\langle m_q|\hat{U}^{(0)\,\dag}(t')\,\hat{H}^{(1)}(t')\,|n_p \rangle^{(0)}\,h^2\nonumber\\ 
&+\frac{i}{\hbar}\,\int_0^tdt'\,e^{-i\omega_qm_q t'}\,e^{i\omega_pn_p t}\,e^{-i\omega_pn_p t'}\,{}^{(0)}\langle m_q|\hat{H}^{(2)}(t')\,|n_p \rangle^{(0)}\,h^2\,\nonumber\\
&-\frac{1}{\hbar^2}\,\int_0^t\int_0^{t'}dt'dt''\,e^{-i\omega_qm_q t'}\,e^{i\omega_pn_p (t-t'')}\,{}^{(0)}\langle m_q|\,\hat{H}^{(1)}(t')\,\hat{U}^{(0)}(t') \hat{U}^{(0)\,\dag}(t'')\,\hat{H}^{(1)}(t'')\,|n_p\rangle^{(0)}\,h^2.
\end{align}
Where we have employed the identity ${}^{(0)}\langle m_q|n_p\rangle^{(0)}=\delta_{m_q n_p}\,\delta_{p q}$. At this point we can make a useful observation which will reduce the number of calculations necessary. Looking at the terms in the first and second moment of the characteristic function, equations \eqref{equ:1moment} and \eqref{equ:2moment} respectively, the second order conditional probability distribution is multiplied by the term $(E_{m_q}^{f(0)}-E_{n_p}^i)$. This term will vanish whenever $m_q=n_p$. Upon finding the squared modulus of the above equation we can see that the majority of second order terms will be derived from the second order terms in equation \eqref{equ:unitary thing} being multiplied by the zeroth order term containing $\delta_{m_q,n_p}$ thereby never contributing to the moments of the characteristic function. The only term which will eventually contribute is the first order term in equation \eqref{equ:unitary thing} multiplied by its modulus. Substituting $\hat{H}^{(1)}$ into this term
\begin{align}
	&\frac{i}{\hbar}\,\int_0^tdt'\,e^{-i\omega_qm_q t'}\,e^{i\omega_pn_p (t-t')}\,{}^{(0)}\bra{m_q}\hat{H}^{(1)}(t')\,|n_p \rangle^{(0)}\,h\nonumber\\ 
	=&\frac{i}{\hbar}\,\int_0^tdt'\,e^{-i\omega_qm_q t'}\,e^{i\omega_pn_p (t-t')}\big{(}M^{(1)}_{qq}\sqrt{m_q}\sqrt{m_q-1}\delta_{m_q-2,n_p}+N_{qp}\sqrt{m_q}\sqrt{n_p}\delta_{m_q-1,n_p-1}\nonumber\\
	&+\sum_{kk}P^{(1)}_{kk}\delta_{m_q,n_p}+P^{(1)}_{pq}\sqrt{m_q}\sqrt{n_p}\delta_{m_q-1,n_p-1}+M_{pp}^{(1)*}\sqrt{n_p}\sqrt{n_p-1}\delta_{m_q,n_p-2}\big{)}\,h.
\end{align}
We can again throw away the terms which force $m_q=n_p$ and multiplying what's remaining by its complex conjugate. We obtain
\begin{align}\label{equ:second order cond prob}
\frac{1}{\hbar^2}\,\int_0^tdt'\,e^{-i\omega_qm_q t'}\,e^{-i\omega_pn_p t'}\big{(}M^{(1)}_{qq}\sqrt{m_q}\sqrt{m_q-1}\delta_{m_q-2,n_p}+M_{pp}^{(1)*}\sqrt{n_p}\sqrt{n_p-1}\delta_{m_q,n_p-2}\big{)}\nonumber\\ 
\times\int_0^tdt''\,e^{i\omega_qm_q t''}\,e^{i\omega_pn_p t''}\big{(}M^{(1)}_{pp}\sqrt{n_p}\sqrt{n_p-1}\delta_{n_p-2,m_q}+M_{qq}^{(1)*}\sqrt{m_q}\sqrt{m_q-1}\delta_{n_p,m_q-2}\big{)}h^2.
\end{align}
This is the only term of the second order conditional probability that will contribute to the moments of the distribution.

\section{Analysing the work distribution moments} 
In order to describe the first and second moments of the work distribution in terms of the cavity dynamics we need to substitute  $p_{m|n}^{(2)},E_n^{f(1)},E_n^{f(2)}$, equations  \eqref{equ:second order cond prob}, \eqref{equ:first order energy eigenvalue} and \eqref{equ:second order energy eigenvalue} respectively, into the below first and second moments 
\begin{align}
G\left[\psi\right]=&\sum_{n_p} p_n E_{n_p}^{f(1)}h+\sum_{n_p} p_n E_{n_p}^{f(2)}h^2+\sum_{m_q,n_p}p_n p_{m|n}^{(2)}\left(E_{m_q}^{f(0)}-E_{n_p}^i\right)h^2\\
G\left[\psi^2\right]=&\sum_{n_p}p_n\left(E_{n_p}^{f(1)}\right)^2h^2+\sum_{m_q,n_p}p_np_{m|n}^{(2)}\left(E_{m_q}^{f(0)}-E_{n_p}^i\right)^2h^2.
\end{align}
This allows us to obtain the moments of the work distribution for any initial state of the cavity up to second order. 

For completeness we write down the full equations for the first and second moment without specifying either the initial state of the cavity or the physical situation causing its dynamics. We have the first moment
\begin{align}\label{equ:full first moment}
G\left[\psi\right]
=&\sum_{n_p} p_n\left(n_p\,\left(N^{(1)}_{pp}+P^{(1)}_{pp}\right)+\sum_k\,P^{(1)}_{kk}\right)h\nonumber\\ 
&+\sum_{n_p} p_n \Bigg{(}-2\,\frac{2\,n_p+1}{E_{2_p}^{(0)}}|M^{(1)}_{pp}|^2
-4\,\sum_{k\neq p}\frac{n_p+1}{E^{(0)}_{1_p,1_k}}\,|M_{kp}^{(1)}|^2-2\,\sum_{k,l\neq p}\frac{1}{E^{(0)}_{1_k,1_l}}\,|M^{(1)}_{kl}|^2\nonumber\\
&+\sum_{k\neq p}\frac{n_p}{E_{1_p}^{(0)}-E^{(0)}_{1_k}}\,\left|N_{pk}^{(1)}+P_{kp}^{(1)}\right|^2+n_p\,(N^{(2)}_{pp}+P^{(2)}_{pp})+\sum_k\,P^{(2)}_{kk}\Bigg{)}h^2\nonumber\\ 
&+2\,\sum_{n_p}\,(n_p+2)(n_p+1)\,p_n\frac{E_{n_p+2}^{f(0)}-E_{n_p}^i}{\hbar^2}\,\left|\int_0^tdt'\,e^{-2\,i(\omega_pn_p+1) t'}\,M_{pp}^{(1)*}
\right|^2\,h^2,
\end{align}
and the second moment
\begin{align}\label{equ:full var of dist}
G\left[\psi^2\right]
=&\sum_{n_p}p_n\left(n_p\left(N^{(1)}_{pp}+P^{(1)}_{pp}\right)+\sum_k\,P^{(1)}_{kk}\right)^2h^2\nonumber\\ 
&+2\,\sum_{n_p}\,(n_p+2)(n_p+1)\,p_n\frac{E_{n_p+2}^{f(0)}-E_{n_p}^i}{\hbar^2}\,\left|\int_0^t dt'\,e^{-2\,i\omega_p(n_p+1) t'}\,M^{(1)}_{pp}\right|^2\,h^2.
\end{align}
We can immediately make some observation as to the behaviour of these moments under certain dynamics. As  justified in  previous work \cite{Barbado:Baez-Camargo:2018}, cavities with ``free falling'' boundaries under motion or dynamics of background spacetimes are associated with non zero diagonal first order Bogoliubov coefficients. Whereas first order non-diagonal Bogoliubov coefficients are more readily associated with rigid boundary conditions such as those of an impinging gravitational wave \cite{Bruschi:Louko:2013}.  

We can therefore make the following statement up to second order. Only ``free falling'' boundaries under motion or dynamics of background spacetimes contribute to the variance of the work distribution as all the dynamical first order contributions in equation \eqref{equ:full var of dist} are diagonal. Whereas, again to second order, the rigid boundary conditions only affect the mean energy change of the cavity. 

We also remark that considering only a first order expansion of the cavity does not reveal a variance in the work distribution and only a second order treatment of the cavity reveals the terms present in \eqref{equ:full var of dist}. Note also that the terms themselves are composed of first order contributions, indeed the ones that are non zero for ``free falling'' boundaries.  

It is also interesting to see that many of the terms present in the second order energy eigenvalue are actually having a negative effect on the energy change of the cavity. These terms represent other modes in the cavity mixing with the cavity mode under consideration, reducing the cavities energy gain. 
\subsection{Specifying dynamics and initial state}
In order to analyse the work distribution of the cavity we specify the cavities initial state. This choice changes the initial probability of the cavity. Let us start with the cavity initially in the vacuum state, completely devoid of excitations 
\begin{align}
	p_n=|\langle n_p|0 \rangle\langle 0|n_p\rangle|^2=\delta_{n_p,0}.
\end{align}
Recalling that the zeroth order energy eigenvalue is simply $E_n^{(0)}=\hbar\omega_p n_p $ and applying the initial vacuum state reduces the first and second moments of the distribution to the following 
\begin{align}
G\left[\psi\right]
=&\sum_k\,P^{(1)}_{kk}h-\frac{1}{\hbar}\Bigg{(}\frac{1}{\omega_p}\,\left|M^{(1)}_{pp}\right|^2
+4\,\sum_{k\neq p}\frac{1}{\omega_p+\omega_k}\,\left|M_{kp}^{(1)}\right|^2+\sum_{k,l\neq p}\frac{1}{\omega_l+\omega_k}\,\left|M^{(1)}_{kl}\right|^2\Bigg{)}h^2\nonumber\\
&+\sum_k\,P^{(2)}_{kk}h^2
+\frac{4\omega_p}{\hbar}\,\left|\int_0^tdt'\,e^{-2\,i\,\omega_p t'}M^{(1)}_{pp}\right|^2h^2 \nonumber\\
G\left[\psi^2\right]
=&\left(\sum_k\,P^{(1)}_{kk}\right)^2h^2+{8\omega_p^2}\,\left|\int_0^tdt'\,e^{-2\,i\,\omega_p t'}M^{(1)}_{pp}\right|^2\,h^2.\label{equ:var vacuum}
\end{align}
In order to analyse the work distribution of the cavity we must also specify the Bogoliubov coefficients responsible for the perturbation of the systems spacetime. We choose those found in equation \eqref{bogoliubov:coefficient} which model an incoming gravitational wave on a 3-dimensional BEC confined in a cavity, neglecting the vacuum contributions to the work distribution present in the terms containing $P^{(1)}_{kk}$. The plot of the variance of the work distribution following this infringement of a gravitational wave can be found in Fig  \ref{fig:protocol} of the main text. 

What can immediately be seen is that rate at which the variance grows is far greater when there is a resonance between the gravitational wave and the confined BEC modes. We can also see that the variance of the work distribution oscillates with a set frequency, which from an inspection of \eqref{equ:var vacuum} is dependant upon the frequency of the cavity mode.  A practical implication of this effect being that in order to extract an amount of work with more certainty, one should conduct their work extraction procedure to coincide with a dip in this variance, which in tern can be tuned dependant upon the cavity parameters. 
\section{Main results for arbitrary initial states}\label{appendix:two}
We finally specialise to the regimes considered in our work. In particular, the transformation $\boldsymbol{S}(t)$ will represent time evolution and therefore we identify $\lambda\equiv t$. Furthermore, in agreement with standard time evolution, we expect $\boldsymbol{S}^{(0)}=\text{diag}(\exp[-i\,\boldsymbol{\omega}\,t],\exp[i\,\boldsymbol{\omega}\,t])$ to be a diagonal matrix such that $\boldsymbol{S}^{(0)}\,\boldsymbol{S}^{(0)\dag}=\mathds{1}$. This implies also that $T^{(0)}_n(t)=T$. Furthermore, the initial state $\boldsymbol{\sigma}_{\text{S}}(0)$ simply coincides with the reduced state of the system S obtained from the full initial state $\boldsymbol{\sigma}_{\text{i}}$. Finally, the frequencies $\omega_n$ have the simple expression $\omega_n=\frac{\pi\,n\,c}{L}$.

\subsection{Main results: changes in energy and entropy}
We proceed to compute the perturbative changes in energy and entropy. These are then used to obtain the efficiency for different choices of initial states.

\subsubsection{First order changes in energy and entropy}
We start with the first order contributions to energy and entropy.
These contributions are defined by the first order contributions $\sigma_{\text{S},nm}^{(1)}$ to the reduced covariance matrix of the system S, which read
\begin{align}\label{important:first:order:cm:elements}
U_{\text{S},nm}^{(1)}=&\sum_{p\in\text{C}}\,\left[\alpha^{(0)*}_{mm}\,\alpha^{(1)}_{np}\,U_{\text{i},pm}+\alpha^{(0)}_{nn}\,\alpha^{(1)*}_{mp}\,U_{\text{i},np}+\alpha^{(0)}_{nn}\,\beta^{(1)*}_{mp}\,V_{\text{i},np}+\alpha^{(0)*}_{mm}\,\beta^{(1)}_{np}\,V_{\text{i},pm}^*\right]\nonumber\\
V_{\text{S},nm}^{(1)}=&\sum_{p\in\text{C}}\,\left[\alpha^{(0)}_{mm}\,\beta^{(1)}_{np}\,U_{\text{i},pm}+\alpha^{(0)}_{nn}\,\beta^{(1)}_{mp}\,U_{\text{i},np}+\alpha^{(0)}_{nn}\,\alpha^{(1)}_{mp}\,V_{\text{i},np}+\alpha^{(0)}_{mm}\,\alpha^{(1)}_{np}\,V_{\text{i},pm}^*\right]\nonumber\\
U_{\text{S},nn}^{(1)}=&2\,\sum_{p\in\text{C}}\,\Re\left(\alpha^{(0)*}_{nn}\,\alpha^{(1)}_{np}\,U_{\text{i},pn}\right)+2\,\sum_{p\in\text{C}}\,\Re\left(\alpha^{(0)}_{nn}\,\beta^{(1)*}_{np}\,V_{\text{i},np}\right)\nonumber\\
V_{\text{S},nn}^{(1)}=&2\,\sum_{p\in\text{C}}\alpha^{(0)}_{nn}\,\beta^{(1)}_{mp}\,\Re U_{\text{i},np}+2\,\sum_{p\in\text{C}}\,\alpha^{(0)}_{nn}\,\alpha^{(1)}_{np}\,\Re V_{\text{i},np}.
\end{align}
In general, since $\alpha^{(0)}_{nn}=\exp[-i\,\omega_n\,t]$, it follows that the expression \eqref{important:first:order:cm:elements} does not vanish. However, an exception occurs when the first order $\boldsymbol{U}$ matrix and the $\boldsymbol{V}$ matrix are diagonal, and the Bogoliubov coefficients have no diagonal first-order contributions, i.e., $\alpha^{(1)}_{nn}=\beta^{(1)}_{nn}=0$. Such scenarios are not uncommon; states with diagonal $\boldsymbol{U}$ matrix and $\boldsymbol{V}$ matrix are either thermal states or single mode squeezed states, and there have been many studies of realistic applications of cavities confining bosonic fields where the Bogoliubov coefficients have no diagonal first-order contributions \cite{Bruschi:Fuentes:2012,Bruschi:Dragan:2013}.

Note that the case where the $\boldsymbol{U}$ matrix and $\boldsymbol{V}$ matrix are both diagonal but the Bogoliubov coefficients have no off-diagonal first-order contributions is excluded since the only way to obtain this would be a Free Hamiltonian with a parameter-dependent frequency, which we are not going to study in this work.

\subsubsection{Second order changes in energy and entropy}
We now look at the case of second order changes. We have
\begin{align}\label{important:second:order:cm:element}
U_{\text{S},nm}^{(2)}=&\sum_p\left[\alpha^{(0)}_{nn}\,\alpha^{(2)*}_{np}\,U_{\text{i},np}+\alpha^{(0)*}_{mm}\,\alpha^{(2)}_{np}\,U_{\text{i},pm}+\alpha^{(0)}_{nn}\,\beta^{(2)*}_{mp}\,V_{\text{i},mp}+\alpha^{(0)*}_{mm}\,\beta^{(2)}_{np}\,V_{\text{i},pm}\right]\nonumber\\
&+\sum_{pq}\left[\alpha^{(1)}_{np}\,\alpha^{(1)*}_{mq}\,U_{\text{i},pq}+\beta^{(1)}_{np}\,\beta^{(1)*}_{mq}\,U^*_{\text{i},pq}+\alpha^{(1)}_{np}\,\beta^{(1)*}_{mq}\,V_{\text{i},pq}+\alpha^{(1)*}_{mq}\,\beta^{(1)}_{mq}\,V^*_{\text{i},pq}\right].
\end{align}
This term is generally not zero. Together with the first order contribution, we can now proceed to look at the different initial states.

\section{Main results for different classes of initial state}\label{appendix:three}

\subsection{Choices of initial state}
We specialise to relevant initial Gaussian states. We will consider pure thermal states $\boldsymbol{\sigma}_{i,\text{Th}}(T)$, two-mode beam-splitted thermal states $\boldsymbol{\sigma}_{i,\text{BS}}(\theta)$, two single-mode squeezed thermal states $\boldsymbol{\sigma}_{i,\text{SMS}}(s)$ and two-mode squeezed thermal states $\boldsymbol{\sigma}_{i,\text{TMS}}(r)$. Interesting multipartite states of three or more modes could be considered but are left for future work. For the case of two-mode states, we fix the modes to be labelled by $k$ and $\tilde{k}$.
The covariance matrix of the \textit{reduced state} of the modes of interest of the different initial states can be found below. All modes that are not considered are in a thermal state.

\begin{align}\label{initial:states}
&\boldsymbol{\sigma}_{i,\text{Th}}(T)=\text{diag}(\nu_1,\nu_2\,\ldots,\nu_1\,\nu_2\,\ldots),\nonumber\\
&\boldsymbol{\sigma}_{i,\text{BS}}(\theta)=
\begin{pmatrix}
\nu_k\,\cos^2\theta+\nu_{\tilde{k}}\,\sin^2\theta & (\nu_k-\nu_{\tilde{k}})\,\sin\theta\cos\theta & 0 & 0\\
(\nu_k-\nu_{\tilde{k}})\,\sin\theta\cos\theta & \nu_k\,\sin^2\theta+\nu_{\tilde{k}}\,\cos^2\theta & 0 & 0\\
0 & 0 & \nu_k\,\cos^2\theta+\nu_{\tilde{k}}\,\sin^2\theta & (\nu_k-\nu_{\tilde{k}})\,\sin\theta\cos\theta\\
0 & 0 & (\nu_k-\nu_{\tilde{k}})\,\sin\theta\cos\theta & \nu_k\,\sin^2\theta+\nu_{\tilde{k}}\,\cos^2\theta
\end{pmatrix},\nonumber\\
&\boldsymbol{\sigma}_{i,\text{SMS}}(s)=
\begin{pmatrix}
\nu_k\,\cosh (2\,s) & 0 & \nu_k\,\sinh (2\,s) & 0\\
0 & \nu_{\tilde{k}}\,\cosh (2\,s) & 0 & \nu_{\tilde{k}}\,\sinh (2\,s)\\
\nu_k\,\sinh (2\,s) & 0 & \nu_k\,\cosh (2\,s) & 0\\
0 & \nu_{\tilde{k}}\,\sinh (2\,s) & 0 & \nu_{\tilde{k}}\,\cosh (2\,s)
\end{pmatrix},\nonumber\\
&\boldsymbol{\sigma}_{i,\text{TMS}}(r)=
\begin{pmatrix}
\nu_k+(\nu_k+\nu_{\tilde{k}})\sinh^2 r & 0 & 0 & \frac{1}{2}(\nu_k+\nu_{\tilde{k}})\sinh (2\,r)\\
0 & \nu_{\tilde{k}}+(\nu_k+\nu_{\tilde{k}})\sinh^2 r & \frac{1}{2}(\nu_k+\nu_{\tilde{k}})\sinh (2\,r) & 0\\
0& \frac{1}{2}(\nu_k+\nu_{\tilde{k}})\sinh (2\,r) & \nu_k+(\nu_k+\nu_{\tilde{k}})\sinh^2 r & 0\\
\frac{1}{2}(\nu_k+\nu_{\tilde{k}})\sinh (2\,r) & 0 & 0 & \nu_{\tilde{k}}+(\nu_k+\nu_{\tilde{k}})\sinh^2 r
\end{pmatrix}.
\end{align}

\subsection{Main results for different classes of initial state without first order diagonal contributions}
Here we assume that $\alpha_{kk}^{(1)}=\beta_{kk}^{(1)}=0$. This is the case for the class of transformations covered by the general techniques described in \cite{Bruschi:Louko:2013}.

We note that all the results presented here are valid for an initial \textit{finite} temperature $T$.

\subsubsection{Initial thermal state}\label{appendix:two:one:passive}
In this case the initial state is $\boldsymbol{\sigma}_{i,\text{Th}}(T)$ and its expression can be found in \eqref{initial:states}. Its is not difficult to see that, in this case, $\nu_n^{(1)}=\sigma_{\text{S},nn}^{(1)}=0$ for all $n$. Therefore, we need to look at the second order contributions to the energy $E_\text{S}$. After some algebra we find
\begin{align}\label{perturbative:efficiency:thermal state:first:case:appendix}
\frac{1}{\eta}=&1+\frac{\sum_{k_\text{E}\in E}\sum_{p\in\text{C}}\,k_\text{E}\,\mathcal{Z}_{k_\text{E}p}(t)}{\sum_{k\in S}\sum_{p\in\text{C}}\,k\,\mathcal{Z}_{kp}(t)},
\end{align}
where we have introduced the function 
\begin{align}
\mathcal{Z}_{nm}(t):=|\beta^{(1)}_{nm}|^2\,(\nu_n+\nu_m)+|\alpha^{(1)}_{nm}|^2\,(\nu_n-\nu_m).
\end{align}
This function will appear also later.

\subsubsection{Initial beam splitted thermal state}\label{appendix:two:one:passive:two}
In this case we have that the system S is composed of two modes with labels $k,\tilde{k}$. These results can be easily generalised to more modes. This implies that the initial state has the property that $U_{\text{i},kk}\neq0$, $U_{\text{i},\tilde{k}\tilde{k}}\neq0$ and $U_{\text{i},k\tilde{k}}\neq0$, while $V_{\text{i},nm}=0$ for all $n,m$, see \eqref{initial:states}. Therefore, we have
\begin{align}\label{important:first:order:cm:elements:two:one:passive:two}
\sigma_{\text{S},kk}^{(1)}=2\,\Re\left(\alpha^{(0)*}_{kk}\,\alpha^{(1)}_{k\tilde{k}}\,U_{\text{i},k\tilde{k}}\right), \quad\quad \sigma_{\text{S},\tilde{k}\tilde{k}}^{(1)}=2\,\Re\left(\alpha^{(0)*}_{\tilde{k}\tilde{k}}\,\alpha^{(1)}_{\tilde{k}k}\,U_{\text{i},\tilde{k}k}\right).
\end{align}
 This allows us to easily find
\begin{align}\label{perturbative:efficiency:squeezed:thermal state:second:case:appendix}
\frac{1}{\eta}=&1+\frac{\sum_{k_\text{E}\in E}\sum_{p\in\text{C}}\,k_\text{E}\,\mathcal{Z}_{k_\text{E}p}(t)}{(k-\tilde{k})(\nu_k-\nu_{\tilde{k}})\,\sin(2\,\theta)\Re\left(\alpha_{kk}^{(0)}\,\alpha_{k\tilde{k}}^{(1)*}\right)}\,h.
\end{align}
Notice that $0\leq\theta<\pi/2$ and that $\theta\gg h$ in our calculations above. For the cases where $\theta\rightarrow0$ we have to proceed differently. Doing so allows us to recover the same result as of the thermal state, which is expected.

\subsubsection{Initial single-mode squeezed thermal state}\label{appendix:two:one:singlemodesqueezed}
In this case we have that the system S is composed of at least of two modes $k,\tilde{k}$. An initial thermal two-mode squeezed state of modes $k,\tilde{k}$ with squeezing parameter $r$ has the following elements: $U_{\text{i},kk}=\nu_k\,\cosh^2 r+\nu_{\tilde{k}}\,\sinh^2 r$, $U_{\text{i},\tilde{k}\tilde{k}}=\nu_{\tilde{k}}\,\cosh^2 r+\nu_k\,\sinh^2 r$ and $V_{\text{i},k\tilde{k}}=V_{\text{i},\tilde{k}k}=(\nu_k+\nu_{\tilde{k}})\,\sinh r\,\cosh r$. In addition, $U_{\text{i},pp}=\nu_p$ for $p\neq k,\tilde{k}$. All together implies
\begin{align}\label{energy:expression:covariance:matrix:appendix:two:mode:squeezed:state}
\frac{1}{\eta}=&1+\frac{P_\text{E}(t)}{P_\text{S}(t)},
\end{align}
where the functions $P_\text{E}(t)$ and $P_\text{S}(t)$ are defined as
\begin{align}\label{p:functions:appendix}
P_\text{E}(t):=&\sum_{k_\text{E}\in E}\sum_{p\in\text{C}}\,k_\text{E}\,\mathcal{Z}_{k_\text{E}p}(t)+k_\text{E}\,\Re\left(\alpha_{k_\text{E}k}^{(1)}\,\beta_{k_\text{E}k}^{(1)*}\right)\,\nu_k\,\sinh(2\,s)+k_\text{E}\,\Re\left(\alpha_{k_\text{E}\tilde{k}}^{(1)}\,\beta_{k_\text{E}\tilde{k}}^{(1)*}\right)\,\nu_{\tilde{k}}\,\sinh(2\,s)\nonumber\\
&+k_\text{E}\,|\beta_{k_\text{E}k}^{(1)}|^2(\nu_k\,\cosh(2\,s)+\nu_{k_\text{E}})+k_\text{E}\,|\beta_{k_\text{E}\tilde{k}}^{(1)}|^2(\nu_{\tilde{k}}\,\cosh(2\,s)+\nu_{k_\text{E}})+k_\text{E}\,|\alpha_{k_\text{E}k}^{(1)}|^2(\nu_k\,\cosh(2\,s)-\nu_{k_\text{E}})\nonumber\\
&+k_\text{E}\,|\alpha_{k_\text{E}\tilde{k}}^{(1)}|^2(\nu_{\tilde{k}}\,\cosh(2\,s)-\nu_{k_\text{E}})\nonumber\\
P_\text{S}(t):=&2\,k\,\nu_{\tilde{k}}\,\sinh(2\,s)\,\Re\left(\alpha_{kk}^{(0)}\,\beta_{kk}^{(2)*}\right)+2\,k\,\nu_{\tilde{k}}\,\sinh(2\,s)\,\Re\left(\alpha_{k\tilde{k}}^{(0)}\,\beta_{k\tilde{k}}^{(2)*}\right)+k\,\sum_{p\in E}\,|\beta^{(1)}_{kp}|^2\,(\nu_p+\nu_k\,\cosh(2\,s))\nonumber\\
&+k\,\sum_{p\in E}\,|\alpha^{(1)}_{kp}|^2\,(\nu_p-\nu_k\,\cosh(2\,s))+|\beta^{(1)}_{k\tilde{k}}|^2\,(\nu_{\tilde{k}}+\nu_k)\,\cosh(2\,s)+|\alpha^{(1)}_{k\tilde{k}}|^2\,(\nu_{\tilde{k}}-\nu_k)\,\cosh(2\,s).
\end{align}

\subsubsection{Initial two-mode squeezed thermal state}\label{appendix:two:one:twomodesqueezed}
In this case we have that the system S is composed of at least of two modes $k,\tilde{k}$. For simplicity we assume that it is composed of two modes only, without loss of generality. These results can be easily generalised to more modes. An initial thermal two-mode squeezed state of modes $k,\tilde{k}$ with squeezing parameter $r$ has the following elements: $U_{\text{i},kk}=\nu_k\,\cosh^2 r+\nu_{\tilde{k}}\,\sinh^2 r$, $U_{\text{i},\tilde{k}\tilde{k}}=\nu_{\tilde{k}}\,\cosh^2 r+\nu_k\,\sinh^2 r$ and $V_{\text{i},k\tilde{k}}=V_{\text{i},\tilde{k}k}=(\nu_k+\nu_{\tilde{k}})\,\sinh r\,\cosh r$. In addition, $U_{\text{i},pp}=\nu_p$ for $p\neq k,\tilde{k}$. All together implies
\begin{align}\label{energy:expression:covariance:matrix:appendix:two:mode:squeezed:state}
\frac{1}{\eta}=&1+\frac{\sum_{k_E\in E}\,k_\text{E}\,U^{(2)}_{f,k_\text{E}k_\text{E}}}{(k+\tilde{k}) (\nu_k+\nu_{\tilde{k}}) \Re\left(\alpha_{kk}^{(0)}\,\beta_{k\tilde{k}}^{(1)*}\right) \sinh(2\,r)} h,
\end{align}
where the function $U^{(2)}_{f,k_\text{E}k_\text{E}}$ is defined as
\begin{align}\label{U2:function:appendix}
U^{(2)}_{f,k_\text{E}k_\text{E}}:=&|\beta_{k_\text{E}k}^{(1)}|^2(\nu_k\,\cosh(2\,r)+\nu_{k_\text{E}})+|\beta_{k_\text{E}\tilde{k}}^{(1)}|^2(\nu_{\tilde{k}}\,\cosh(2\,r)+\nu_{k_\text{E}})\nonumber\\
&+|\alpha_{k_\text{E}k}^{(1)}|^2(\nu_k\,\cosh(2\,r)-\nu_{k_\text{E}})+|\alpha_{k_\text{E}\tilde{k}}^{(1)}|^2(\nu_{\tilde{k}}\,\cosh(2\,r)-\nu_{k_\text{E}})\nonumber\\
&+(\nu_k+\nu_{\tilde{k}})\,\sinh(2\,r)\,\Re\left(\alpha_{k_\text{E}k}^{(1)}\,\beta_{k_\text{E}\tilde{k}}^{(1)*}\right)+(\nu_k+\nu_{\tilde{k}})\,\sinh(2\,r)\,\Re\left(\alpha_{k_\text{E}\tilde{k}}^{(1)}\,\beta_{k_\text{E}k}^{(1)*}\right)+\sum_{p\in E}\mathcal{Z}_{k_\text{E}p}(t).
\end{align}

\subsection{Main results for different classes of initial state with first order diagonal contributions}
Here we assume that $\alpha_{kk}^{(1)}=0$ and $\beta_{kk}^{(1)}\neq0$. This is the case for the class of transformations covered by the general techniques described in \cite{Barbado:Baez-Camargo:2018}.

In this case, we have that the first order corrections to the energy are sufficient, and read
\begin{align}
U_{S,nn}=2\,\sum_{p\in\text{C}}\,\Re\left(\alpha^{(0)*}_{nn}\,\alpha^{(1)}_{np}\,U_{\text{i},pn}\right)+2\,\sum_{p\in\text{C}}\,\Re\left(\alpha^{(0)}_{nn}\,\beta^{(1)*}_{np}\,V_{\text{i},np}\right).
\end{align}
These are the only terms we are interested in. We follow the procedure outlined above and obtain the efficiency for the cases of interest here.

\subsubsection{Initial thermal state}\label{appendix:two:two:thermal}
We find, for the case of the initial purely thermal state
\begin{align}\label{perturbative:efficiency:thermal state:second:case:appendix}
\frac{1}{\eta}=&1+\frac{\sum_{k_\text{E}\in E}\,k_\text{E}\,\Re\left(\alpha^{(0)*}_{k_\text{E}k_\text{E}}\,\alpha^{(1)}_{k_\text{E}k_\text{E}}\right)\,\nu_{k_\text{E}}}{\sum_{k\in S}\,k\,\Re\left(\alpha^{(0)*}_{kk}\,\alpha^{(1)}_{kk}\right)\,\nu_k}.
\end{align}

\subsubsection{Initial thermal beam-splitted state}\label{appendix:two:two:thermal:beam:splitted}
Moving on to the initial thermal beam-splitted state we find
\begin{align}\label{energy:expression:covariance:matrix:appendix:active}
\frac{1}{\eta}=&1+\frac{\sum_{k_\text{E}\in E}\,k_\text{E}\,\Re\left(\alpha^{(0)*}_{k_\text{E}k_\text{E}}\,\alpha^{(1)}_{k_\text{E}k_\text{E}}\right)\,\nu_{k_\text{E}}}{R(t)},
\end{align}
where we have defined
\begin{align}
R(t)=&k\,\Re\left(\alpha^{(0)*}_{kk}\,\alpha^{(1)}_{kk}\right)\,\nu_k+2\,k_{\tilde{k}}\,\Re\left(\alpha^{(0)*}_{\tilde{k}\tilde{k}}\,\alpha^{(1)}_{\tilde{k}\tilde{k}}\right)\,\nu_{\tilde{k}}
+(k-\tilde{k})(\nu_k-\nu_{\tilde{k}})\Re\left(\alpha^{(0)*}_{kk}\,\alpha^{(1)}_{k\tilde{k}}\right)\,\sin\theta\,\cos\theta\nonumber\\
&-(\nu_k-\nu_{\tilde{k}})\,\left(k\,\Re\left(\alpha^{(0)*}_{kk}\,\alpha^{(1)}_{kk}\right)\,\nu_k-\tilde{k}\,\Re\left(\alpha^{(0)*}_{\tilde{k}\tilde{k}}\,\alpha^{(1)}_{\tilde{k}\tilde{k}}\right)\,\nu_{\tilde{k}}\right)\,\sin^2\theta
\end{align}
for convenience of presentation.

\subsubsection{Initial single mode squeezed thermal state}\label{appendix:two:two:squeezed}
Finally, we focus on a single mode squeezed thermal state of mode $k$, that has the form
\begin{align}
\boldsymbol{\sigma}_{i,\text{SMS}}(s)=
\nu_k\,\begin{pmatrix}
\cosh (2\,s) & \sinh (2\,s)\\
\sinh (2\,s) & \cosh (2\,s)
\end{pmatrix}.
\end{align}
We therefore find
\begin{align}\label{energy:expression:covariance:matrix:appendix:squeezed}
\frac{1}{\eta}=&1+\frac{\sum_{k_\text{E}\in E}\,k_\text{E}\,U^{(2)}_{f,k_\text{E}k_\text{E}}}{2\,\nu_k\,\Re\left(\alpha^{(0)*}_{kk}\,\beta^{(1)}_{kk}\right)\,\sinh(2\,s)}\,h,
\end{align}
where we have defined 
{\small
\begin{align}
U^{(2)}_{f,k_\text{E}k_\text{E}}:=&|\beta_{k_\text{E}k}^{(1)}|^2(\nu_k\,\cosh(2\,s)+\nu_{k_\text{E}})+|\alpha_{k_\text{E}k}^{(1)}|^2(\nu_k\,\cosh(2\,s)-\nu_{k_\text{E}})+\nu_k\sinh(2\,s)\,\Re\left(\alpha_{k_\text{E}k}^{(1)}\,\beta_{k_\text{E}k}^{(1)*}\right)+\sum_{p\in E}\mathcal{Z}_{k_\text{E}p}(t)
\end{align}
}

\end{document}